\documentclass[a4paper,fleqn,usenatbib]{mnras}

\usepackage{ae,aecompl}
\usepackage{graphicx}	% Including figure files
\usepackage{amsmath}	% Advanced maths commands
\usepackage{amssymb}	% Extra maths symbols
\usepackage{subfigure}

\title[Hot Jupiter - Angular Momentum and Heat Transport]{Angular Momentum and Heat Transport on Tidally Locked Hot Jupiter Planets}

\author[J. M. Mendon\c ca]{
Jo\~ao M. Mendon\c ca$^{1,2}$\thanks{E-mail: joao.mendonca@space.dtu.dk}\thanks{Homepage: software-oasis.com}
\\
$^{1}$National Space Institute, Technical University of Denmark, Elektrovej, 2800, Kgs. Lyngby, Denmark.\\
$^{2}$Department of Physics, University of Oxford, Clarendon Laboratory, Parks Road, Oxford, U.K.\\
}

\date{Accepted XXX. Received YYY; in original form ZZZ}

\pubyear{2019}

\begin{document}
\label{firstpage}
\pagerange{\pageref{firstpage}--\pageref{lastpage}}
\maketitle

% Abstract of the paper
\begin{abstract}
The atmospheric circulation in the upper atmosphere of hot Jupiter planets is strongly influenced by the incoming stellar radiation. In this work we explore the results from a 3D atmospheric model and revisit the main processes driving the circulation in hot Jupiter planets. We use the angular momentum transport as a diagnostic and carry out a Fourier analysis to identify the atmospheric waves involved. We find that the coupling between the angular momentum transported horizontally by the semi-diurnal tide and the mean circulation is the mechanism responsible for producing the strong jet at low latitudes.

Our simulations indicate the possible formation of atmospheric indirect cells at low latitudes. The formation of these cells is induced by the presence of the semi-diurnal tide that is driven by the stellar irradiation. The tropical circulation has an important impact transporting heat and momentum from the upper towards the lower atmosphere. One of the consequences of this heat and momentum transport is a global increase of the temperature. 

We show that the initial conditions do not affect the output of the reference simulation. However, when the period of rotation of the planet was increased ($P_{rot} > 5$ Earth days), vertical transport by stationary waves became stronger, transient waves became non-negligible, and Coriolis influence less dominant, which allowed a steady state with a strong retrograde jet to be stable. We found that at least two statically steady state solutions exist for the same planet parameters.\\

\end{abstract}


%%%%%%%%%%%%%%%%%%%%%%%%%%%%%%%%%%%%%%%%%%%%%%%%%%

%%%%%%%%%%%%%%%%% BODY OF PAPER %%%%%%%%%%%%%%%%%%

\section{Introduction}
\subsection{Background}
Hot Jupiter planets are a class of giant gaseous planets with no comparable example in our solar system in terms of, for example, extreme atmospheric thermal forcing due to their proximity to the parent star. The thermal structure in these astronomical objects is largely affected by the heat transport (see the review article by \citealt{2015Heng}). The response of the atmosphere to radiative forcing varies as a function of pressure. The radiative timescale can change from roughly 8 hours at 100 hPa to 2.3 Earth days at 1000 hPa (\citealt{2005Iro}). This timescale increases very rapidly for higher pressures, where it becomes larger than the sidereal day of the planet. At those pressures, the tidal effects become negligible and the thermal structure is mostly driven by the flow motion. At lower pressures, the radiative timescale decreases and the day-night contrasts increase. The 3D temperature structures formed in the atmosphere are a balance of this radiative timescales and dynamical timescales, which depend on the magnitude of the flow motion. To be able to understand the mechanisms that are driving the atmospheric circulation, a good estimate of these two quantities is important. 

Hot Jupiter planets inward of 0.05 AU are expected to be tidally locked due to the large tidal stresses induced by the parent star (e.g., \citealt{1996Guillot} and \citealt{2014Showman}). The rotation periods of the giant gaseous planets in our solar system varies from the short 10 hours for Jupiter to 17 hours for Uranus. These values are shorter than the ones estimated for a typical hot Jupiter planet which is in general larger than 2 Earth days (e.g., \citealt{2007Udry}). This difference in the rotation rate implies that the rotational effects in a hot Jupiter atmosphere are less efficient driving the atmosphere. This phenomenon can be quantified by a non-dimensional parameter called the thermal Rossby number, which is the ratio between the inertial and Coriolis forces. From phase curve observations of the hot Jupiter HD 189733b, evidence of strong winds that shift the maximum flux in the secondary eclipse are expected to be present (\citealt{2009Knutson} and \citealt{2012Knutson}), which suggests a thermal Rossby number of the order of unity (e.g., \citealt{2011ShowmanCho}) and larger than any giant gaseous planet in our solar system ($<0.1$). In \cite{2011ShowmanCho} it is also estimated that the Rossby deformation radius, which is the length scale at which the rotation effects become as important as the buoyancy effects, is close to the planetary radius. This implies that large-scale eddies are expected to be present in these planets, including atmospheric cells that can extend from the equator to the pole. There are other parameters that are often used to describe the global circulation such as the Rhines scale. This number can estimate the number of multiple zonal jets from geostrophic turbulence (see \citealt{1975Rhines}). The scale is related to the latitudinal extent of parallel baroclinic zones, and the ratio of the planetary radius over this scale can be associated with the number of jets. The Rhines scale is also a quantity which measures the influence of the rotational effects in the atmospheric flow. In \cite{2011ShowmanCho}, the number of jets was estimated to be between 1 and 5 in hot Jupiter planets, which depend largely on the magnitude of the flow. These numbers are rough estimates. A formulation of a similarity theory which describes the flow in hot Jupiter planets from a set of characteristic dimensionless parameters will require larger effort from theory, numerical simulations and observations of these astronomical objects.

\subsection{Theoretical motivation}
One of the main features in the atmospheric circulation of tidally locked hot Jupiter planets obtained with 3D simulations is the strong equatorial jet (e.g., \citealt{2002Showman}; \citealt{2005Cooper}; \citealt{2009Menou};\citealt{2010Rauscher}; \citealt{2009Showman}; \citealt{2011Heng}; \citealt{2013Dobbs-Dixon}; \citealt{2013Parmentier}; \citealt{2014Mayne}; \citealt{2015Kataria}; \citealt{2016Mendoncab}). This jet has an important contribution in the heat transport from the day to night side. In \cite{2011Showman} the mechanisms that form this jet were studied. The formation of such jet at low latitudes requires angular momentum transport in the up-gradient direction (\citealt{1969Hide}). It is proposed by \cite{2011Showman} that global-scale stationary waves\footnote{Stationary waves are waves at rest in the reference frame. In this study the reference frame is moving with the planet rotation and the substellar point (tidally locked planet), which means that the thermal tides are stationary waves. Transient waves are not at rest in the reference frame.}, produced by the strong heating in the day side due to the irradiation from the parent star (thermal tides), transport angular momentum horizontally towards low latitudes and have an important contribution in producing the strong prograde jet at the equator. These global-scale waves transporting angular momentum were identified as a combination of equatorial Rossby and Kelvin waves. It is interesting to note that in this case the vertical transport by the stationary waves is associated with the deceleration of the strong jet (\citealt{2011Showman}), which is in contrast with the results of 3D simulations on Venus (e.g. \citealt{2010Lebonnois}; \citealt{2016Mendonca}). Venus is the second closest terrestrial planet in our solar system (0.7 AU) with a massive atmosphere (it is roughly $92\times10^4$ hPa at the surface) covered by very opaque clouds and its atmospheric circulation is also driven by global-scale waves excited by the periodic forcing of the stellar radiation (e.g. \citealt{2016Mendonca}). In Venus, numerical models and observations indicate the presence of a strong equatorial jet in the cloud region (e.g.,\citealt{1985Kerzhanovich}; \citealt{2008Lavega}; \citealt{2009Moissl}; \citealt{2012Mendonca}), where most of the incoming radiation from the sun is absorbed. In this case the transport of the upward propagating waves excited in the cloud region is the main mechanism accelerating the low latitude jet. More theoretical discussion about the angular momentum transport can be found in section \ref{sec:transpmom}. Titan is another case in the Solar System that presents an atmosphere in the superrotating state. However, the mechanisms driving the fast winds in Titan's atmosphere are different from the ones in Venus or hot Jupiter planets. The main difference is associated with the long radiative adjustment timescale in Titan that weakens the impact of the thermal tides driving the circulation. 3D numerical simulations of Titan (e.g., \citealt{2011Newman}; \citealt{2012Lebonnoisa}) show that the formation of strong winds at low latitudes are associated with formation of synoptic waves which in turn formed due to the presence of barotropic unstable high-latitude jets that transport angular momentum against the transport due to the mean meridional circulation.

Using 3D atmospheric simulations \cite{2017Mayne} argued that the formation of the equatorial jet in hot Jupiter planets is a combination of eddy momentum flux and mean circulation. This theory is consistent with previous works of \cite{2011Showman} and \cite{2014Tsai}. The work presented here, extends previous studies by exploring how the angular momentum is transported by mean circulation and waves obtained from Global Circulation Model (GCM) simulations. The main goal is to have a better knowledge on how the angular momentum and heat are being redistributed across the atmosphere of hot Jupiter planets. This will allow to have a more complete understanding on the main mechanisms driving the circulation. The techniques we use quantify the transport of angular momentum by mean circulation, atmospheric stationary and transient waves, and allow us to verify directly which process is dominant.  The methods used here are complementary to the Eliassen-Palm diagnostic fluxes analysed in \cite{2017Mayne}, which provide direct measure of the locations in the atmosphere where the mean flow is being accelerated by mean circulation and waves (i.e., regions of convergence of angular momentum fluxes). In our work, we applied a Fourier analysis to the wind and temperature fields to determine the properties of the atmospheric waves present in the simulations. By analysing the atmospheric waves we want to determine which waves have a larger impact driving the atmosphere and assess if the waves excited in the simulated atmospheres can allow multiple equilibria solutions of the atmospheric circulation. In this work, we also want to explore how the heat is being redistributed across the atmosphere to search for atmospheric processes that can lead to the warm-up of the global temperature structure and be associated with the ``inflated'' hot Jupiters phenomenon.

In section \ref{sec:model}, we describe the model used, followed by section \ref{sec:basesimu} where the results of a defined reference simulation are analyzed. In section \ref{sec:transpmom}, the angular momentum transported by the mean circulation and waves is studied. The tools used in this section are important to study the mechanisms that drive the atmospheric circulation in planetary atmospheres. In section \ref{sec:waveanalisis}, the waves responsible for transporting angular momentum are identified. In section \ref{sec:MES}, we explore cases of multiple solutions for the atmospheric circulation. Finally, concluding remarks are presented in section \ref{sec:conclu}.

\section{Model}
\label{sec:model}
In this work a comprehensive GCM was used to explore the atmospheric circulation of giant gaseous planets under extreme levels of irradiation, such as hot Jupiter planets. This model includes a quasi-hydrostatic dynamical core, a simplified radiative transfer scheme and a convection scheme that mixes potential temperature and momentum under super-adiabatic conditions.

In Table \ref{tab:model} the general model parameters, which represent the parent star and the planet studied are shown. These are approximate numbers that represent typical hot Jupiter planets, and do not try to exemplify any specific system. Below, we describe the dynamical and physical packages, which are included in the GCM.

\begin{table}
\begin{center}
\caption{Model parameters used in the baseline simulation.}
\begin{tabular}{ | l | l | l |}
\hline
 Parameters &  & Units  \\ \hline \hline
 Star Temperature & 4500 & K \\ \hline
 Star Mass        & 0.67 & Solar Mass \\ \hline
 Planet distance & 0.05 & AU \\ \hline
 Mean Radius & 70000 & km \\ \hline
 Gravity & 10 & m/s$^2$ \\ \hline
 Gas constant & 3500 & J/K/kg \\ \hline
 Specific heat & 13000 & J/K/kg \\ \hline
 Highest pressure & 100000 & hPa \\ \hline
 Sidereal orbit period & 5 & Earth days \\ \hline
 Sidereal rotation period & 5 & Earth days \\ \hline
 Orbit inclination & 0 & deg \\ \hline
 Orbit eccentricity & 0 & deg \\ \hline
 Constant upward heat flux \\ (lowest pressure model interface) & 120 & kW/m$^2$\\ \hline
 \label{tab:model}
 \end{tabular}
\end{center}
\end{table} 

\subsection{Dynamical Core}
\label{subsec:dcore}
The dynamical core solves the equation that represents the dynamical behaviour of the resolved global atmospheric flow. For this work we used the dynamical core version 4.5.1 from the UK Meteorological Office Portable Unified Model (UM). This model has been used with success in numerical Earth weather and climate research (e.g., \citealt{2008Reichler}; \citealt{2000Stott}), and also for atmospheric dynamical studies of Venus and Jupiter (e.g., \citealt{2004Yamazaki} and \citealt{2016Mendonca}).

The nonlinear differential fluid equations are solved using a finite-difference formulation. The equations included were derived in \cite{1995White}. The approximations used in these equations  are less drastic than the common ones used for the meteorological primitive equations such as neglecting the metric and the cosine Coriolis terms (``traditional approximation'') and/or the shallow atmosphere approximation. For more information see \cite{1995White} and \cite{1992Cullen}.

The equations in the dynamical core are solved over a latitude-longitude grid.  This horizontal discretization of space is associated with the convergence of the meridians at the poles which largely constrains the time-step at high latitudes to maintain the model stability. In order to relax the Courant-Friedrichs-Lewy (CFL) condition at high latitudes, the dynamical core uses a longitudinal filtering of fast waves and flow (Fourier filter). 

A horizontal numerical diffusion is also included to remove numerical noise and qualitatively represents the physical phenomena of turbulence and eddy viscosity on the sub-grid scale. The strength of the diffusion was chosen to be as low as possible while keeping the model numerically stable: $2.7\times10^{11}$(dimensionless quantity, \citealt{1992Cullen}) increasing slightly at the top of the model domain and the order applied was 6th. The horizontal and vertical space resolutions are $5^\circ\times5^\circ$ for the horizontal and 37 layers for the vertical, which covers the atmosphere from roughly $100\times10^3$ hPa up to a pressure level of roughly 0.035 hPa. For the reference simulation presented below, tests with twice the horizontal resolution did not change the main results discussed in this work. The time-step for the dynamical core was 300 seconds.

\subsection{Physics Core} 
\label{subsec:phys}
Coupled with the dynamical core we included two parameterizations which represent the radiative processes and convection in the atmosphere. We also implemented two ``sponges'' that damp the eddy component of the wind (\citealt{2018Mendoncab}), one at the top and other at the bottom of the model domain to prevent any unphysical reflections with the boundaries of the ``rigid lids'' of our model atmosphere. The time-step in the physical core was set to 600 seconds.

\subsubsection{Radiative transfer scheme}
\label{subsec:rtm}
The radiation scheme implemented is based on an absorptivity/emissivity formulation (neglecting scattering processes). The spectral integration is divided in two parts (two bands): one that solves the incoming absorption of the stellar radiation, and other which represents the thermal emission/absorption of the atmosphere (including also a prescribed internal heat flux at the bottom). This simple scheme allowed us to explore easily different possible thermal structures from close-in hot Jupiter planets (see the Appendix in \cite{2018Mendoncaa} for more information on the scheme). 

\begin{figure}
\begin{center}
\vspace{0.1in}
\includegraphics[width=0.9\columnwidth]{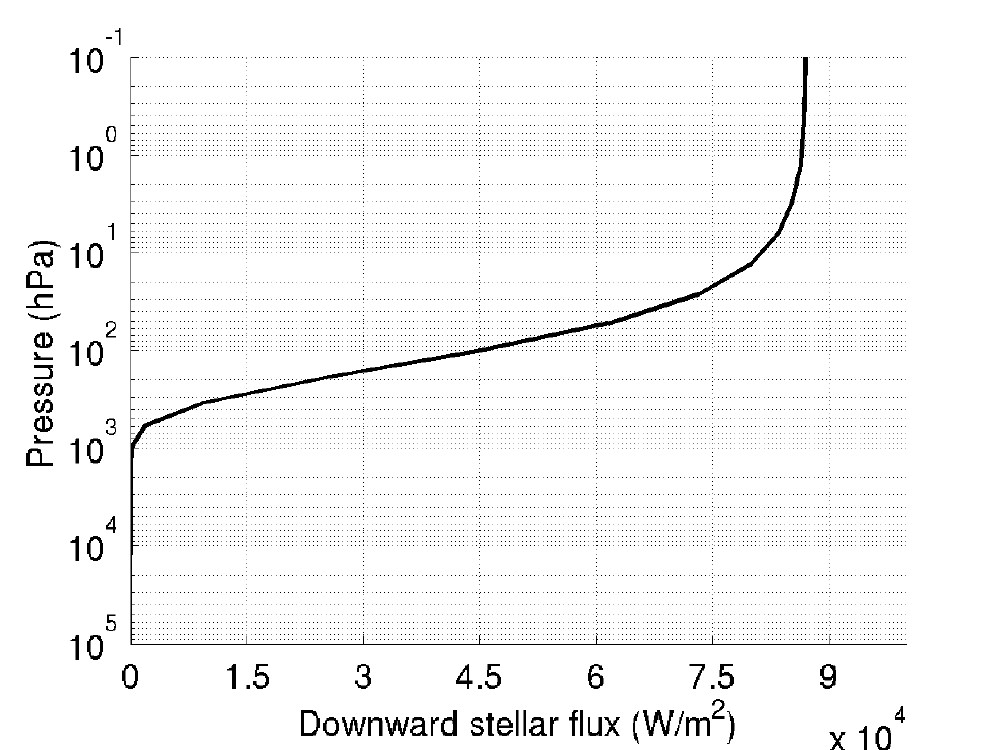}
\end{center}
%\vspace{-0.2in}
\caption{Global average downward stellar flux computed from a 1D radiative transfer code. Note: 1 hPa is equivalent to 1 mba.}
%\vspace{0.1in}
\label{fig:fdsg}
\end{figure}

\begin{figure}
\begin{center}
\vspace{0.1in}
\includegraphics[width=0.9\columnwidth]{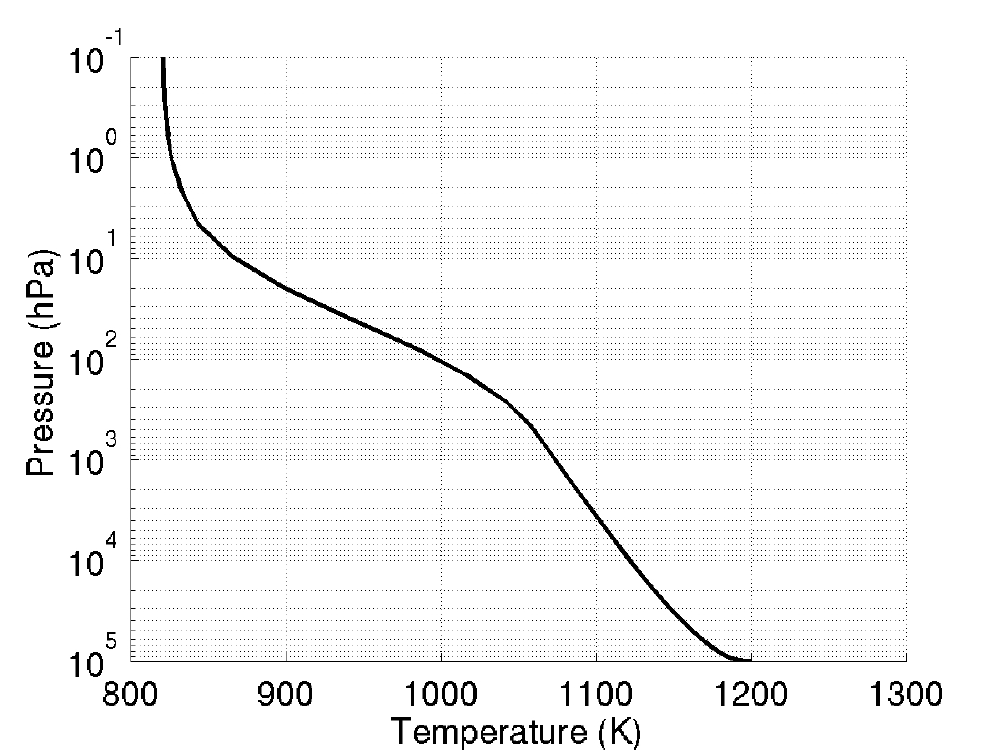}
\end{center}
%\vspace{-0.2in}
\caption{Vertical temperature profile in radiative-convective equilibrium obtained by a 1D model for reference values $\tau_0$ and $\tau_1$.}
%\vspace{0.1in}
\label{fig:1DTemps}
\end{figure}

We define the optical depth in the atmosphere with the following form,
\begin{equation}
\label{eq:tausw}
\tau_{stellar}= \frac{\tau_{0}}{p_{\star}}(p_{bot}-p_{top}), 
\end{equation} 
\begin{equation}
\tau_{thermal}= \frac{\tau_{1}}{p_{\star}}(p_{bot}-p_{top} + \frac{1}{p_{\star}}(p_{bot}-p_{top})^2),
\end{equation}
where $p_{\star}$ is the pressure of the model's bottom domain (100 bar), and $p_{bot}$ and $p_{top}$ are the pressure at the bottom and top of the atmospheric layers. These are two simple equations that represent the atmospheric opacity and are controlled by the tunable parameters: $\tau_{0}$ and $\tau_{1} $. The physical quantities in each layer are assumed to be horizontally homogeneous. In this work we explore the case $\tau_0/\tau_1 = 1.0$.  The value of $\tau_{stellar}$ was chosen so that the downward stellar flux would drop to roughly 0 at 1000 hPa (see Fig. \ref{fig:fdsg}), and set the photosphere to roughly 100 hPa. These numbers were based on results shown in e.g., \citealt{2016Read}. Different values of $\tau_{stellar}$ allow the distribution of stellar energy be deposited in different regions of the atmosphere. In Fig. \ref{fig:1DTemps} we show the radiative-convective equilibrium temperature profile obtained by using the reference $\tau_0$ and $\tau_1$. The 1D model couples the simple radiative transfer and the convection scheme described below. The heating rates produced by the stellar radiation were globally averaged using an 8-point Gaussian quadrature method, so that the final temperature profile can be related to a global averaged profile.

\subsubsection{Convection}
\label{subsubsec:conv}
A simple convective adjustment scheme was included. This parameterization represents the vertical mixing of potential temperature ($\theta$) and momentum in a buoyantly unstable atmospheric column. The instability condition is satisfied when $\delta \theta/\delta p > 0$. The total enthalpy is conserved during the vertical mixing. The final mean potential temperature is found assuming a dry atmosphere,
\begin{equation}
\label{equ:mixpt}
\bar{\theta} =\frac{\int^{p_{top}}_{p_{bot}} \theta \Pi d p}{\int^{p_{top}}_{p_{bot}} \Pi dp}
\end{equation}
where $\Pi$ is the exener function ($(p/p_0)^{Rd/Cp}$), and $p_{top}$ and $p_{bot}$ are the pressures at the top and bottom of the unstable column. We also take into account the mixing of momentum in the unstable atmospheric column. The intensity of the momentum mixing is difficult to estimate even with more complex parameterizations, and is usually constrained using empirical parameters. However, we use a simple approximation, which was first suggested by \cite{1993Hourdin}. The intensity of the momentum mixing depends on the amplitude of the thermal adjustment:
\begin{equation}
\nonumber           
\alpha = \frac{\int_{p_{bot}}^{p_{top}} |\theta - \bar{\theta}| dp}{\int_{p_{bot}}^{p_{top}} \bar{\theta} dp}, \;\bar{u} = \frac{\int_{p_{bot}}^{p_{top}} u dp}{\int_{p_{bot}}^{p_{top}} dp}, \;\bar{v} = \frac{\int_{p_{bot}}^{p_{top}} vdp}{\int_{p_{bot}}^{p_{top}} dp},  
\end{equation} 
\begin{equation}
u_{new} = u + \alpha(\bar{u} - u), 
\end{equation} 
\begin{equation}
\nonumber    
 v_{new} = v + \alpha(\bar{v} - v), 
\end{equation}
where $u$ is the wind speed in the longitudinal direction and $v$ in the meridional direction, and $\alpha$ is the weight of the momentum mixing. $u_{new}$ and $v_{new}$ are the new wind speed values after the mixing. The momentum mixing has a larger impact for larger instabilities, and $\bar{\theta}$ is the mean potential temperature computed in equation \ref{equ:mixpt}.

\subsubsection{Boundary Layer}
\label{subsubsec:BL}
At the three top layers and at the bottom ($10\times10^5$ hPa) of the model's domain, sponge parameterizations are used to damp the \emph{eddy} component of the wind field \emph{only} to zero. These sponges avoid any unphysical wave reflections at the model's boundaries.

\section{Reference simulation}
\label{sec:basesimu}
\begin{figure}
\begin{center}
\vspace{0.1in}
\includegraphics[width=0.9\columnwidth]{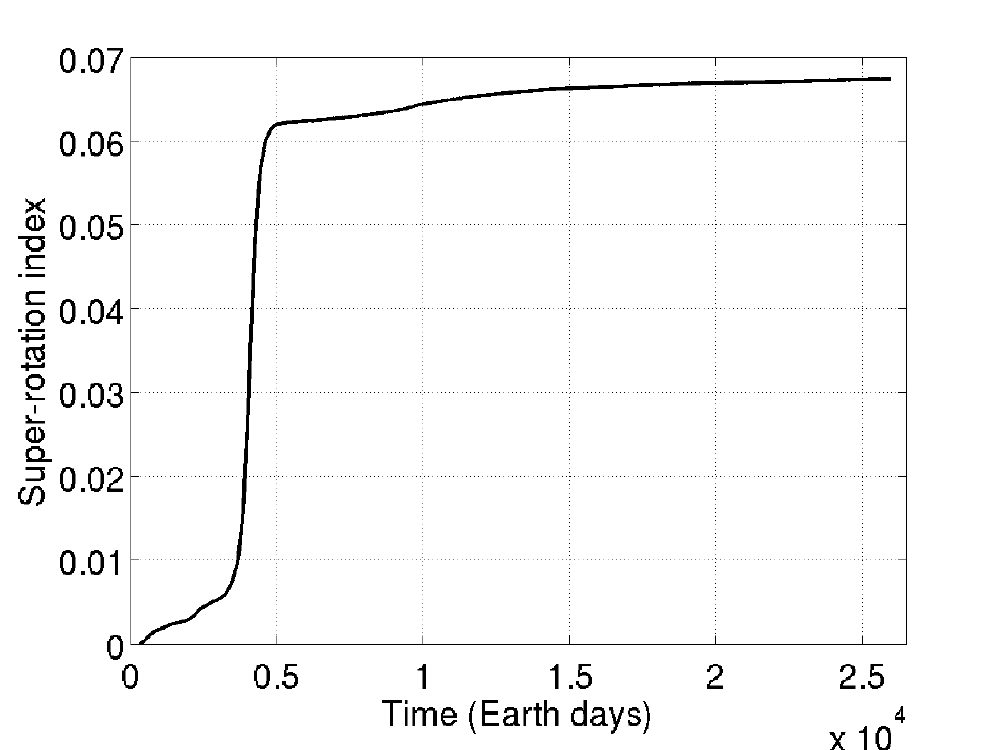}
\end{center}
%\vspace{-0.2in}
\caption{Global super-rotation index for the reference simulation.}
%\vspace{0.1in}
\label{fig:SRind}
\end{figure}

The reference simulation started from a rest state and an isothermal atmosphere (1000 K). The stellar energy deposited at different regions in the atmosphere plus an upward energy flux coming from the deep atmosphere, coupled with the instabilities due to the planet's rotation, start the dynamical forcing in the atmosphere. Available potential energy is converted to kinetic energy. In this work, the statistical equilibrium state is found when the total axial angular momentum of the atmosphere stays roughly at the same level as a function of time. Before this period, the model is in the ``spin-up phase''. In our experiments, this period has a duration of 26500 Earth days which is larger than previous studies (e.g., \citealt{2002Showman}; \citealt{2005Cooper}; \citealt{2009Showman}; \citealt{2010Thrastarson}; \citealt{2011Heng}; \citealt{2017Mayne}; \citealt{2018Mendoncaa}). However, a long period such as the one estimated in this work is expected due to the large thermal inertia of the deep atmosphere. Fig. \ref{fig:SRind} shows the total axial angular momentum of the atmosphere as a function of time over the total axial angular momentum of the same atmosphere at the rest state minus one. This quantity is also known as the global super-rotation index or the Read number (\citealt{1986Read}). This parameter is often used to classify if atmospheres are super-rotating or not, and has been shown to be higher for slowly rotating planets such as Venus and Titan (e.g., \citealt{1995Hourdin}, \citealt{2016Mendonca}). As shown in Fig. \ref{fig:SRind}, the global super-rotation index in the simulation increases with time during the spin-up phase. The variations in the global super-rotation index are associated with the non conservative angular momentum processes included in the GCM: convection, explicit diffusion, intrinsic diffusion from the numerical methods in the dynamical core, filters and the sponge layers. The non conservative processes are not a disadvantage in the model since their weights are much smaller than the actual physical exchanges of angular momentum in the atmosphere studied here. The global super-rotation index is a mass weighted number so the deepest regions have the largest contributions. In the reference simulation, the conservative exchanges of momentum in the atmosphere bring retrograde flow to the deepest layers.  Retrograde means that the atmosphere is rotating in the opposite direction as the considered rotation of the planet. When the dissipative terms act on the retrograde flow they drive an increase in the total axial angular momentum. The opposite would happen if prograde flow was present in the deepest layers. The same trend is seen in other simulations of hot Jupiter planets using a different dynamical core (\texttt{THOR}, \citealt{2016Mendoncab}). The jump at 5000 Earth days in Fig. \ref{fig:SRind} is related to a dynamical adjustment in the deep atmosphere (below the pressure level $10^4$ hPa). During the spin-up phase the model tries to converge to its more stable regime and during this period non-physical sources associated with the numerical methods and boundary conditions can have a non-negligible impact on how the physical variables evolve. The sudden jump would have been harder to track if, for example, the total kinetic energy of the atmosphere was being used as the prognostic for the spin-up phase, because the magnitude of the winds were kept roughly the same but the direction changed very quickly. The global super-rotation index converges to roughly 0.07. After 26500 Earth days, the change in the global super-rotation index is very small changing by less than $0.05\%$ in the last 500 Earth days. At this point we assumed that the simulation had reached its statistical steady state.

\begin{figure}
\begin{center}
%\vspace{0.1in}
\includegraphics[width=0.9\columnwidth]{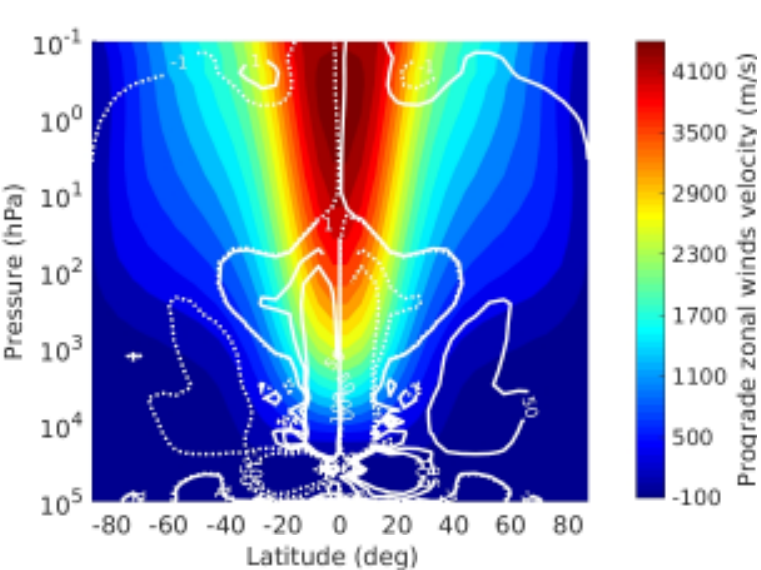}
\end{center}
%\vspace{-0.2in}
\caption{Averaged zonal winds and mass stream function (in units of $10^{10}$kg/s) from the GCM reference simulation. The dashed lines represent the anti-clockwise circulation and the solid lines the clockwise. The values were zonal and time averaged for 100 Earth days.}
%\vspace{0.1in}
\label{fig:baselineu}
\end{figure}

\begin{figure}
\begin{center}
%\vspace{0.1in}
\includegraphics[width=0.9\columnwidth]{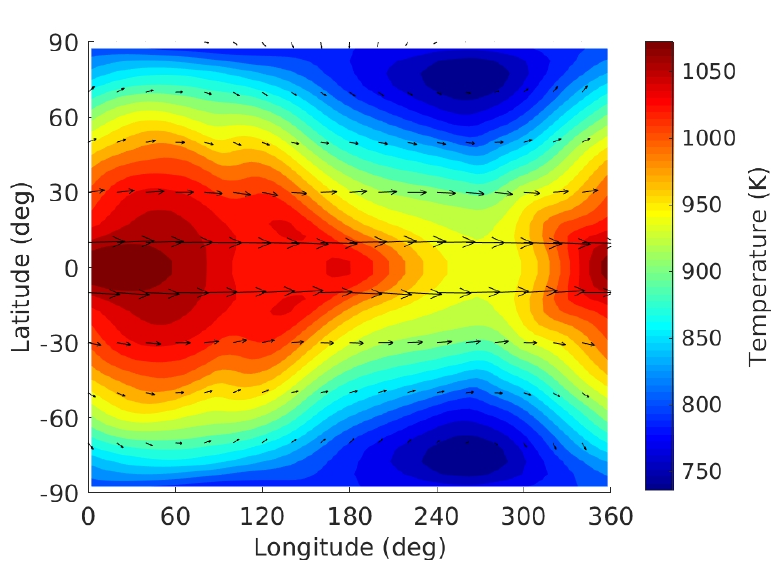}
\end{center}
%\vspace{-0.2in}
\caption{Longitude-latitude map of temperature at 50 hPa. The arrows show the time averaged direction of the wind velocity. The results were averaged over the last 100 Earth days of the simulation. The substellar point is at $0^{o}$ longitude.}
%\vspace{0.1in}
\label{fig:baselineuv_50}
\end{figure}

In Fig. \ref{fig:baselineu}, we show the averaged zonal winds and mass stream function. The color contours show a clear peak in the zonal winds at low latitudes and pressure level of roughly 1 hPa. This equatorial prograde jet reaches values around 4 km/s. The mechanism responsible for the formation of this equatorial maximum at low latitudes is a combination of mean circulation and waves which transport angular momentum up-gradient (e.g., \citealt{2011Showman}, \citealt{2014Tsai} and \citealt{2017Mayne}). This mechanism is identified and explored in section \ref{sec:transpmom}. In the equatorial region the magnitude of the prograde zonal winds decreases with the increase of pressure from the strong jet core until roughly a pressure level of $5\times10^4$ hPa. Near the bottom pressure level there is a region with retrograde zonal winds which extends to higher altitudes for latitudes poleward of 60 deg in each hemisphere. Fig. \ref{fig:baselineuv_50} shows how the temperature is redistributed horizontally at 50 hPa. The highest temperature is shifted eastwards by roughly 15$^{o}$ with respect to the sub-stellar point due to the efficient heat transport by the strong equatorial jet. An infrared light-curve observation would obtain a maximum in emission flux just before the secondary eclipse.
\begin{figure}

\begin{center}
%\vspace{0.1in}
\includegraphics[width=0.9\columnwidth]{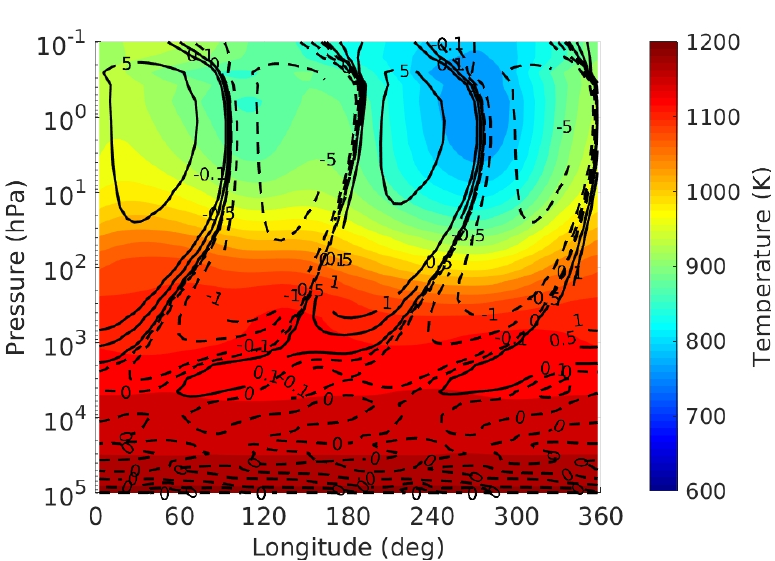}
\end{center}
%\vspace{-0.2in}
\caption{Map of temperature and vertical winds averaged in time and latitude (between -20$^{o}$ and 20$^{o}$ latitude. The colors are the temperatures and the contours are the vertical velocities in m$/$s. The solid lines represent the upward velocities and the dashed lines the downward motion. The results were averaged over the last 100 Earth days of the simulation.}
%\vspace{0.1in}
\label{fig:baselineu_cells}
\end{figure} 

\begin{figure}
\centering
\subfigure[]{\label{fig:vth1}\includegraphics[width=0.4\textwidth]{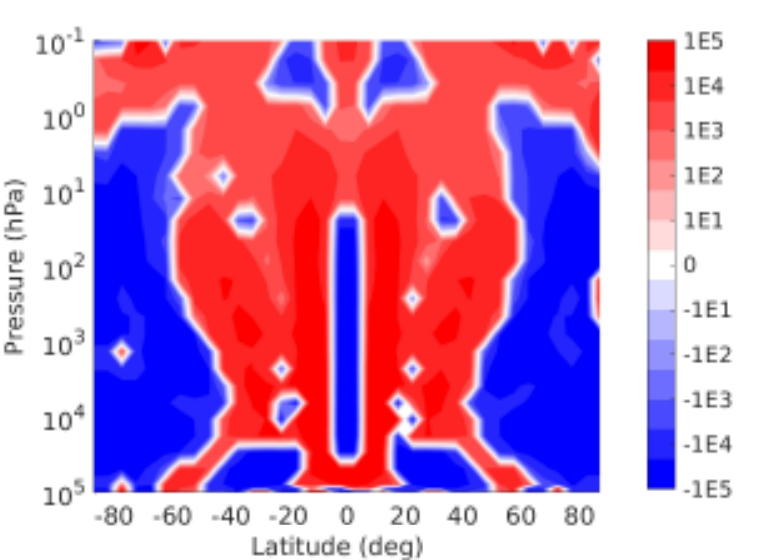}}
\subfigure[]{\label{fig:vth2}\includegraphics[width=0.4\textwidth]{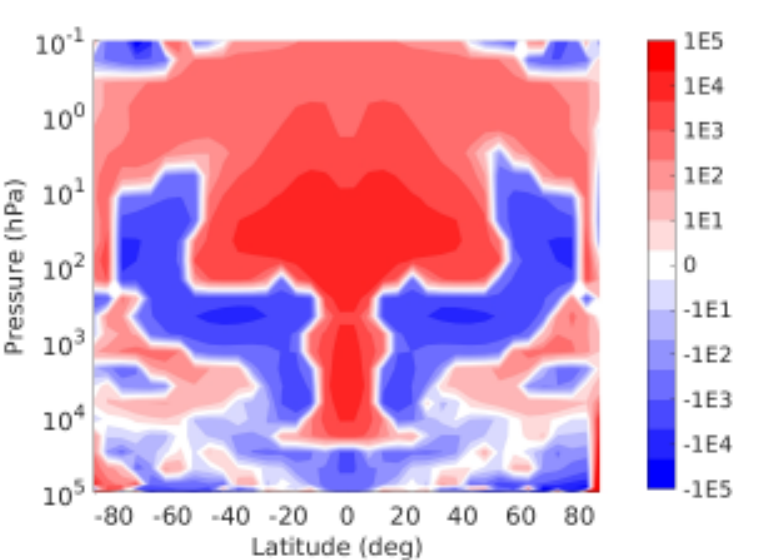}}
\caption{Vertical  heat transport by mean circulation and waves in the reference simulation. The data used to produce these results correspond to the last 100 Earth days of the simulation. \textbf{(a)} is the vertical transport by mean circulation ($[\overline{v}][\overline{T}]$) and \textbf{(b)} by stationary waves ($[\bar{v^{\star}}\bar{T^{\star}}]$).  The units of the colour bars are m s$^{-1}$K.}
\label{fig:vtheat}
\end{figure}

Above 200 hPa, as can be seen in Fig. \ref{fig:baselineu}, a large direct cell drives most of the circulation. A large equator-to-pole circulation in each hemisphere is allowed to form due to the slow planetary rotation, which weakens the Coriolis acceleration and increases the efficiency of the poleward heat transport in the atmosphere. The mass-stream function lines show some complex regions mostly at low latitudes for pressures higher than $1.0\times10^4$ related with the more turbulent flow. Between 50 and $3\times10^4$ hPa the results show indirect\footnote{Also known as eddy driven cells, where air rises over a cold temperature region and sinks over warm a temperature region - e.g., anti-clockwise cells in the northern hemisphere.} cells at low latitudes (similar to the atmospheric cells obtained in \citealt{2015Charnay}). In these indirect cells, hot air is pushed downwards at low latitudes by mean circulation. These atmospheric cells are possible because the eddy momentum convergence at low latitudes dominates over the the atmospheric forcing related to the latitudinal gradient of the radiative heating (see \citealt{2006Vallis} for further discussion on eddy driven atmospheric cells). The essential eddy momentum convergence at low latitudes is exerted by the semi-diurnal tide (second harmonic of the thermal tides). The semi-diurnal tide is excited by the stellar forcing and propagates vertically. Fig. \ref{fig:baselineu_cells}, shows how the vertical velocity is distributed across the equatorial region, and the pattern of the semi-diurnal tide is clearly seen in the contour results with two maxima along the equator. Above 100 hPa the latitudinal gradient of the radiative heating dominates the eddy convergence at low latitudes by absorbing most of the stellar radiation, which enhances the upward motion at low latitudes. However, the radiative heating induced by the stellar flux drops quickly for pressures below 100 hPa and the downward motion becomes more intense. As we see in section \ref{sec:waveanalisis}, the semi-diurnal tide propagates to very deep layers in the atmosphere. In section \ref{sec:waveanalisis} we explore in more detail the properties of the waves present in the simulated atmosphere. The formed indirect cells may have also an important role  vertically mixing chemistry and clouds in the atmosphere. In Fig. \ref{fig:vtheat}, we show maps of the vertical heat transport by mean circulation and stationary waves (stronger than the transient waves as we show in the next section). Map (Fig. \ref{fig:vth1}) shows the heat transport by the mean circulation that clearly dominates the transport by the waves (Fig. \ref{fig:vth2}) in the equatorial region. The efficient transport by the mean circulation at low latitudes is associated with the indirect cells. This atmospheric phenomenon is transporting heat from the upper atmosphere to pressure levels below 10$^4$ hPa. From these results it is not possible to predict if these cells would extend to even deeper regions in the atmosphere if the bottom of the model was set to higher pressures. Deeper cells would be possible if the semi-diurnal tide can propagate to deeper pressures. In our model, a rigid bottom prevents any information being transported below $10^5$ hPa and a constant upward heat flux (a boundary condition for the radiative transfer scheme, see Table \ref{tab:model}) is set to weaken  the amplitude of the waves in the deep atmosphere for numerical stability.

\begin{figure*}
\centering
\subfigure[]{\label{fig:dbtemp1}\includegraphics[width=0.7\columnwidth]{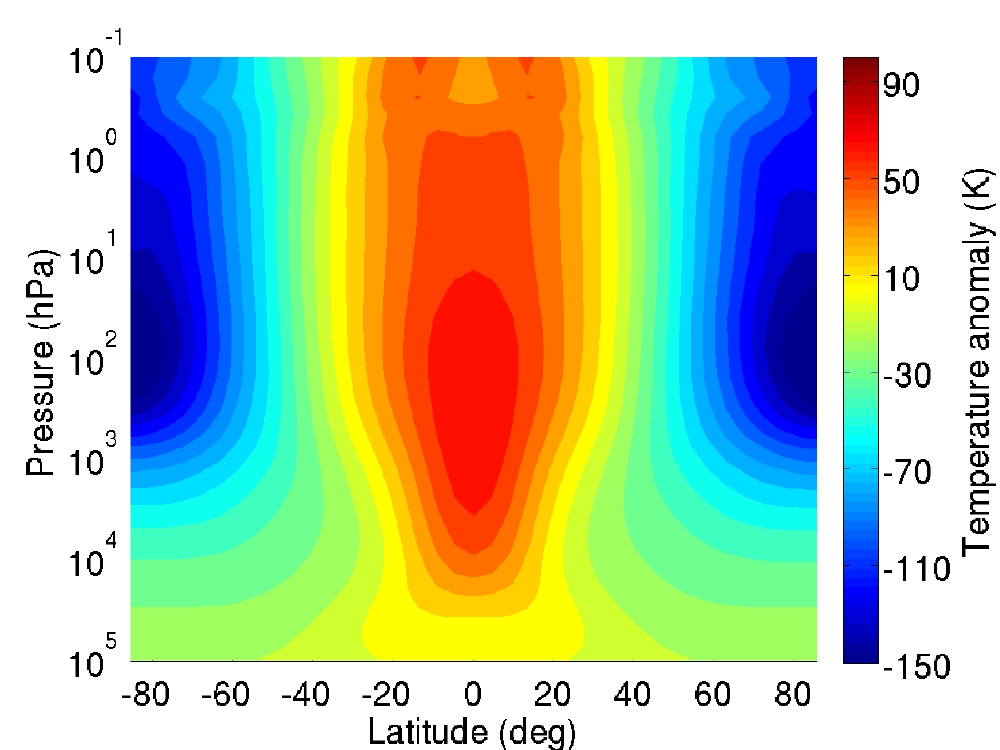}}\hspace{2.0cm}
\subfigure[]{\label{fig:dbtemp2}\includegraphics[width=0.7\columnwidth]{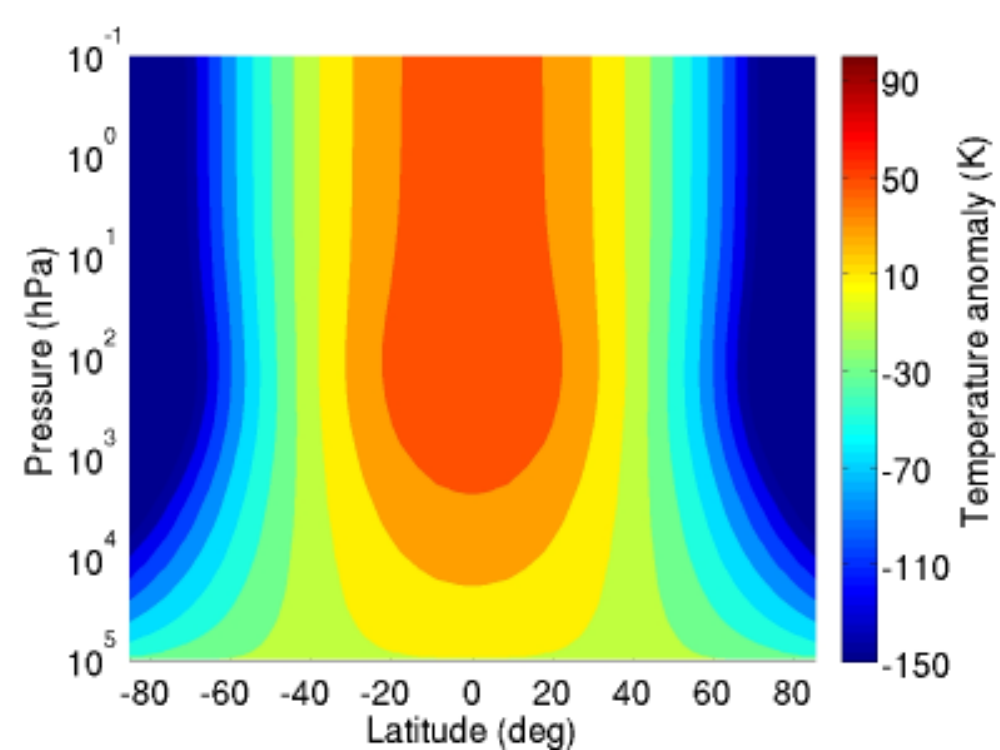}}\\[0.0cm]
\subfigure[]{\label{fig:dbtemp3}\includegraphics[width=0.7\columnwidth]{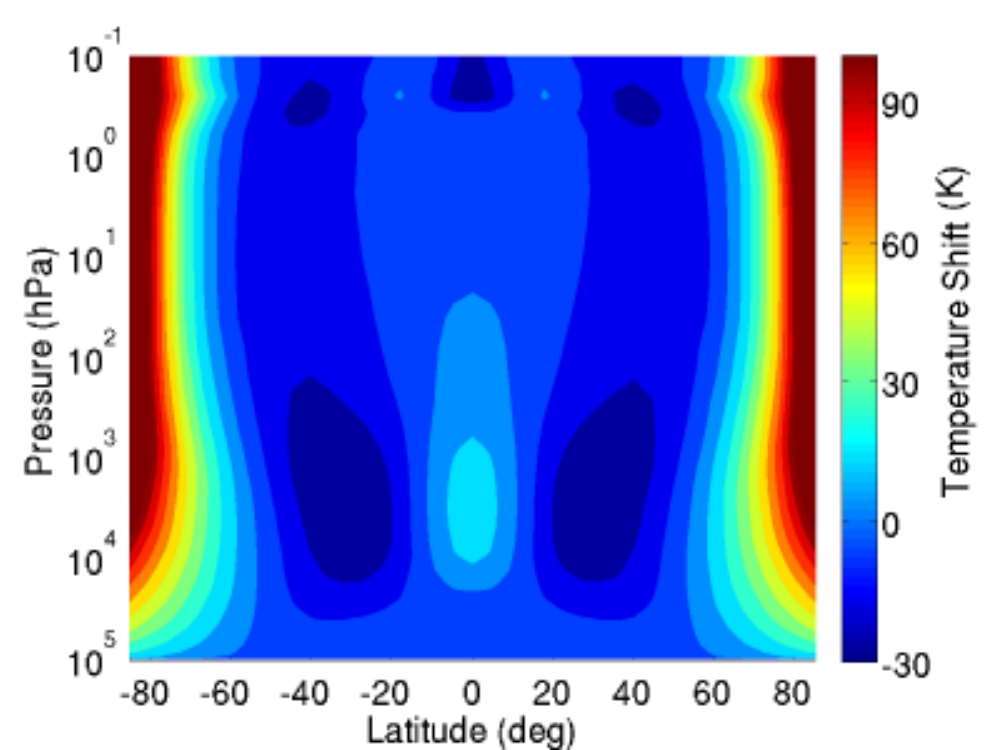}}\hspace{2.0cm}
\subfigure[]{\label{fig:dbtemp4}\includegraphics[width=0.7\columnwidth]{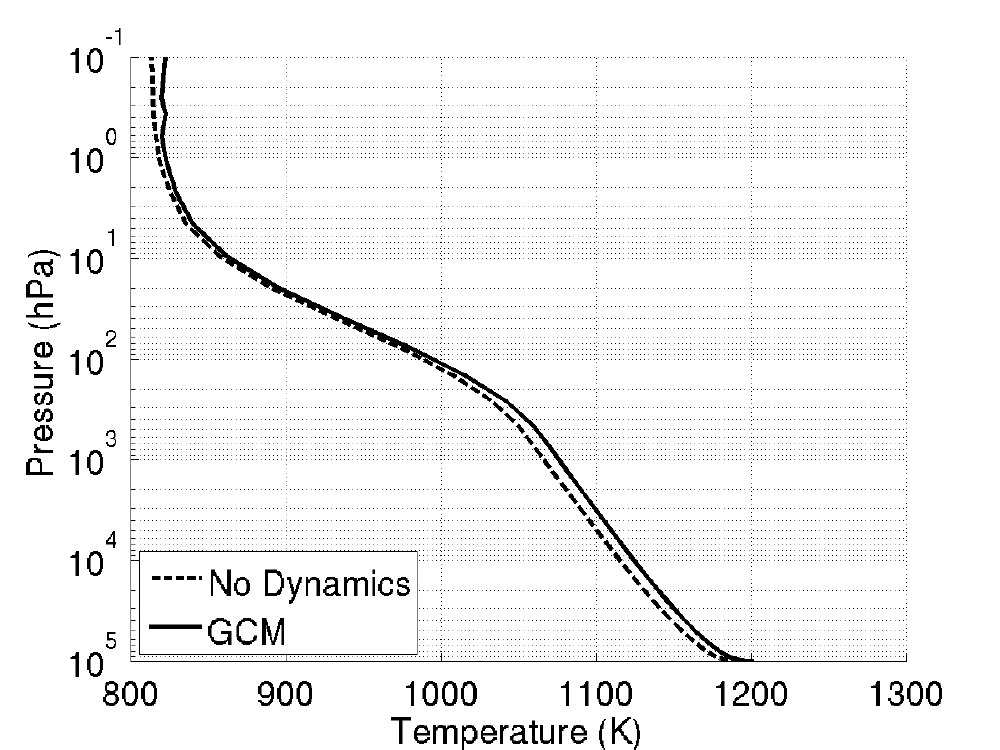}}\\[0.0cm]
\caption{\textbf{(a)} Time (100 Earth days) and zonal averaged temperature anomaly maps obtained at the end of the reference simulation. \textbf{(b)} Temperature anomalies obtained from the zonally averaged radiative equilibrium temperatures at different latitudes computed by the radiative transfer scheme only. The temperature anomalies are computed using equation \ref{eq:tempano}. \textbf{(c)} shows the difference between (a) and (b). \textbf{(d)} Global and time average equilibrium temperatures with (solid line) and without (dashed line) atmospheric dynamics.}
\label{fig:db_temp}
\end{figure*}

Figs. \ref{fig:dbtemp1} and \ref{fig:dbtemp2} show the mean temperature anomaly maps which represent the temperature deviations from the horizontal mean on pressure levels. The temperature anomaly values ($T_a$) were computed from the following equation:
\begin{equation}
\label{eq:tempano}
T_a = T - \frac{\int_{\lambda=0}^{\lambda=2\pi} \int_{\phi=-\pi/2}^{\phi=\pi/2} T a^2\cos\phi d\lambda d\phi}{\int_{\lambda=0}^{\lambda=2\pi} \int_{\phi=-\pi/2}^{\phi=\pi/2} a^2\cos\phi d\lambda d\phi}
\end{equation}
where $T$ is the absolute temperature, $a$ is the radius of the planet, $\phi$ is the latitude and $\lambda$ is the longitude. The deviation in temperature in these figures is influenced by radiative forcing and by the atmospheric circulation. In Fig. \ref{fig:dbtemp1} the temperature contrast between low and high latitudes is clear. The thermal structure in the reference simulation is mainly molded by the deposition of the direct radiation from the parent star in the atmosphere. Neglecting the dynamics in the atmosphere the temperature anomaly map expected would be equal to the one shown in Fig. \ref{fig:dbtemp2}. In the lower atmosphere the main differences are at higher latitudes and as expected due to the larger radiative timescale. When the dynamics is turned on, heat is transported polewards and is trapped in the polar regions. Another possible mechanism to explain hotter temperatures at the polar regions may be compressional adiabatic warming from the descending branch of the large atmospheric cells seen in the mass stream function plot. An interesting phenomenon that is clearly seen in Fig. \ref{fig:dbtemp3} is the extra heat accumulated at low latitudes ($< 10$ hPa) brought by the indirect cells from the upper atmosphere. The heat is forced by the mean flow to be transported downwards at low latitude below 10 hPa, working as a heat pump that mixes potential temperature. Fig. \ref{fig:dbtemp4} shows a lower atmosphere in the simulation with dynamics 10 K warmer than in the simple radiative equilibrium simulation. Despite the warming still being very small to explain the ``inflated'' hot Jupiter planets problem (see \cite{2011Fortney} for a review on possible mechanisms) it may play an important role as the trigger for the observed phenomenon. The downward transport induced by the semi-diurnal tides injects heat in the atmosphere, and there are factors that we do not take into account that may enhance its impact in the temperature structure, if implemented: hotter temperatures in the opacities, a deeper bottom boundary for the model (so cells can extend to even deeper pressures), representation of the interior planet evolution, and conversion from kinetic dissipation (e.g., \citealt{2002Showman}) or MHD (Magnetohydrodynamics, e.g., \citealt{2014Rogers}) drag into heat. We also need to explore the impact of the diffusion. An inappropriately strong diffusion can make fine atmospheric structures such as waves disappear, and consequently prevent the formation of indirect cells. Also, a more detailed analysis on the heat transport and its sensitivity to different stellar/planet and orbit parameters needs to be done in more detail. Observational studies indicate a correlation between the ``inflated'' hot Jupiter planets and the magnitude of the planet's incoming stellar flux (e.g., \citealt{2011Demory}; \citealt{2011Laughlin}; \cite{2011Miller}; \cite{2013Weiss}; \citealt{2014Figueira}). It is important that we develop a deeper understanding of the dependence of the semi-diurnal tides vertical propagation on planet and star parameters, and also explore the formation of possible critical layers that can absorb the waves and prevent their propagation to deeper pressures (e.g., convection regions). A good knowledge of the impact of this phenomenon in the global climate/atmospheric circulation is important to have a better understanding of the coupling between the deep and upper atmosphere. 

\cite{2002Showman} was the first work to suggest the possibility of a downward circulation (described as similar to Walker circulation in the Earth atmosphere) in the atmosphere of a hot Jupiter planet. Later in \cite{2017Tremblin}, it was proposed from two-dimensional steady-state atmospheric circulation model simulations that the anomalously large radii in hot Jupiter planets is associated with the advection of the potential temperature due to mass and longitudinal momentum conservation. As discussed above, in our work we have analysed the complete physical 3D model and discuss the mechanism responsible for forming the important downward flow, show its impact in the 3D temperature structure and suggest the missing ingredients that can enhance its impact in the atmosphere.  

\section{Angular momentum transport}
\label{sec:transpmom}
The strong equatorial jet obtained in the reference simulation cannot be produced or maintained from axisymmetric mechanisms (\citealt{1969Hide}). The well known ``Hides's first theorem'' (\citealt{1986Read}) states that the excess of axial angular momentum can only be obtained from non-axisymmetric motions (e.g., waves). This result can be learned from the following equation of motion:
\begin{equation}
\underbrace{\frac{\partial}{\partial t}(\rho M) }_{[A]}+ \underbrace{\frac{}{}\vec{\nabla}\cdot (\rho \textbf{v}M)}_{[B]} + \underbrace{\frac{\partial p}{\partial \lambda}}_{[C]} = \underbrace{\frac{}{}\vec{\nabla}\cdot(\tau \cdot \hat{\textbf{z}}\times\textbf{r})}_{[D]}
\end{equation}
where $\tau$ is the viscous stress tensor, $\hat{\textbf{z}}$ is the unit vector in the direction of the planetary angular velocity ($\Omega$) and $M$ is the axial angular momentum per unit mass. To simplify this equation we remove the term [C] by zonally averaging the equation and the term [D] by assuming that friction is negligible. The remaining terms represent the conservation of angular momentum under axisymmetric inviscid motion. In the presence of closed circulation such as atmospheric cells this equation cannot explain the presence of local maxima. To produce local maxima in the axial angular momentum per unit of mass in the atmosphere, such as the equatorial jet obtained in the reference results (the equator is the farthest region from the rotation axis), we need convergence of the angular momentum flux, $\textbf{F}_M = \rho \textbf{v} M$, towards that maximum. Non-axisymmetric motions can transport angular momentum towards these regions. This extra zonal pressure torque in the atmosphere can be represented by Reynolds' stress terms:
\begin{equation}
\label{eqn:fm}
\overline{[F_M]} = \rho a \cos \phi (\overline{[u^{\star}v^{\star}]},\overline{[u^{\star}w^{\star}]}) ,
\end{equation}
where $a$ is the planet radius, $\rho$ is the atmospheric density, the square brackets represent zonal averages, the bars over the variables indicate temporal averages, and $v$ and $w$ are the meridional and vertical wind speed respectively. The stars on each variable mean that they are disturbances in relation to their respective zonal average. The two terms in Eq. \ref{eqn:fm} are the eddy flux terms for stationary waves (meaning that they move with the global planet) and can be related with mechanisms which transport angular momentum and form or maintain strong winds at low latitudes for example. The solution for transient waves is very similar, but the disturbances in this case will be in relation to their respective time average (see Eq. \ref{eq:mom_trans}). In order to understand which processes form the strong prograde jet we need to quantify and localize these terms in the atmosphere.

\begin{figure}
\centering
\subfigure[]{\label{fig:dbht1}\includegraphics[width=0.4\textwidth]{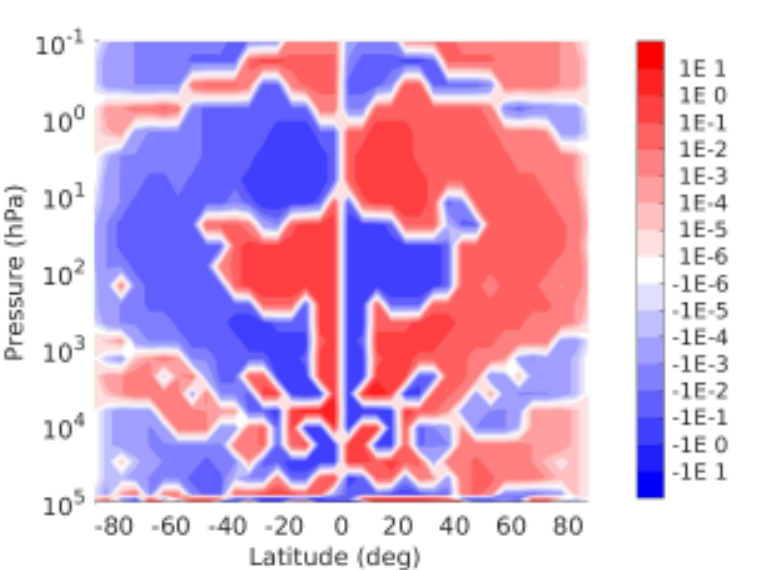}}
\subfigure[]{\label{fig:dbht2}\includegraphics[width=0.4\textwidth]{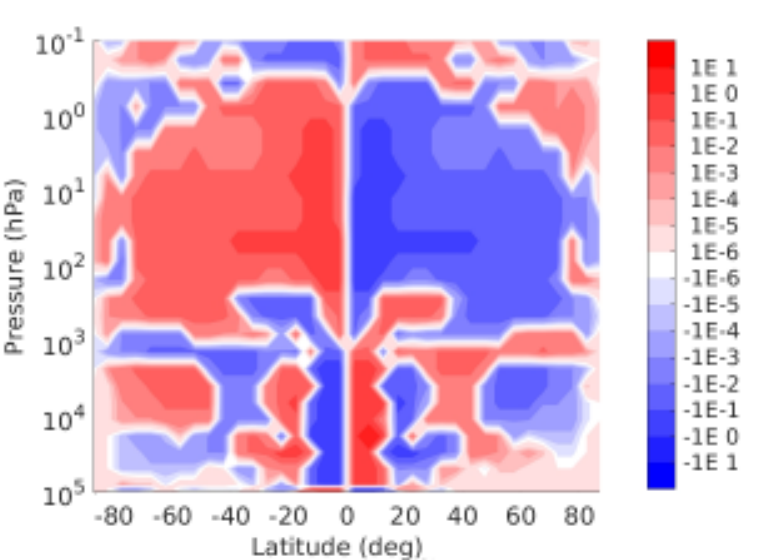}}
\subfigure[]{\label{fig:dbht3}\includegraphics[width=0.4\textwidth]{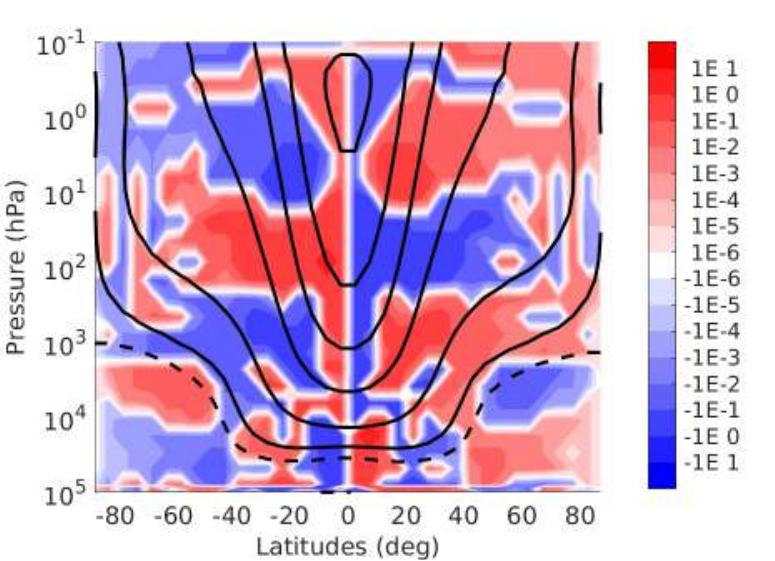}}
\caption{Meridional transport of angular momentum by waves in the reference results. The data used to produce these results correspond to the last 100 Earth days of the simulation. \textbf{(a)} is the horizontal transport by mean ciculation ($[\overline{v}][\overline{M}]$), \textbf{(b)} by stationary waves ($[\bar{v^{\star}}\bar{M^{\star}}]$) and \textbf{(c)} is the net horizontal transport.  The units of the colour bars are $10^{28}$ kg m$^3$ s$^{-2}$. Note that the positive numbers in regions of prograde winds are associated with northward transport of prograde angular momentum, but in regions of retrograde winds positive numbers mean southward transport of retrograde angular momentum (see discussion in section \ref{sec:transpmom}). The solid black lines represent the contours of the averaged zonal winds in m$/$s: 4500, 3500, 2500, 1500 and 500. The dashed line represents the contour of averaged zonal winds equal to zero.}
\label{fig:dbht}
\end{figure}

\begin{figure}
\centering
\subfigure[]{\label{fig:dbvt1}\includegraphics[width=0.4\textwidth]{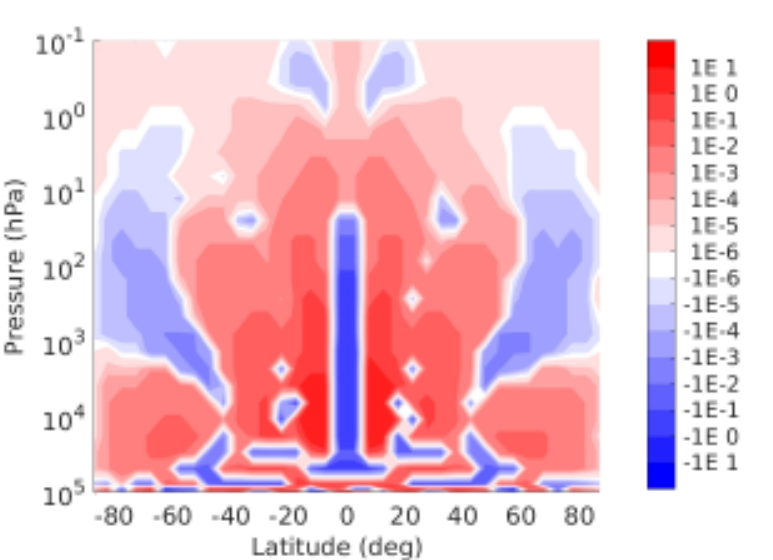}}
\subfigure[]{\label{fig:dbvt2}\includegraphics[width=0.4\textwidth]{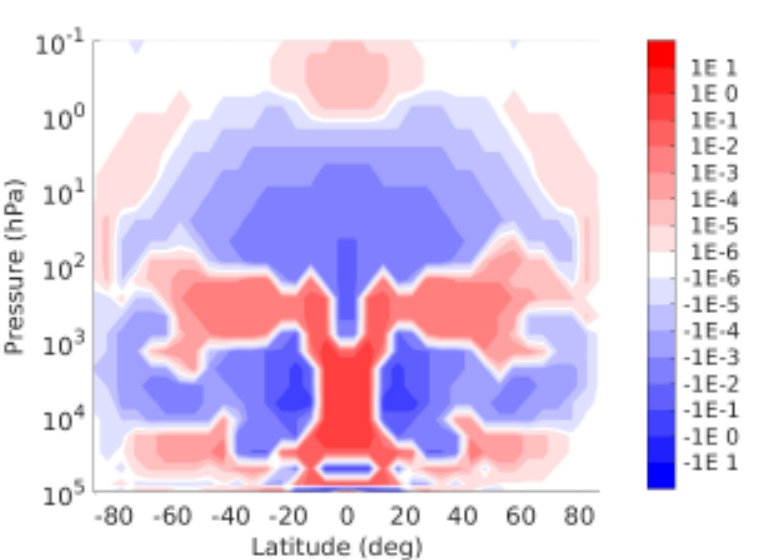}}
\subfigure[]{\label{fig:dbvt3}\includegraphics[width=0.4\textwidth]{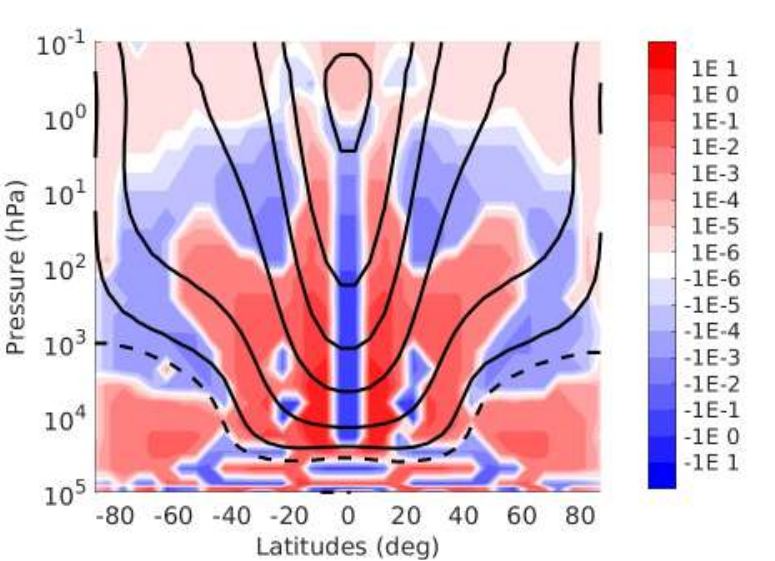}}
\caption{Vertical transport of angular momentum by waves in the reference results. The data used to produce these results correspond to the 100 Earth days of the simulation. \textbf{(a)} is the vertical transport by mean ciculation ($[\overline{w}][\overline{M}]$), \textbf{(b)} by stationary waves ($[\bar{w^{\star}}\bar{M^{\star}}]$) and \textbf{(c)} is the net vertical transport.   The units of the color bars are $10^{31}$ kg m$^3$ s$^{-2}$.  Note that the positive numbers in regions of prograde winds are associated with upward transport of prograde angular momentum, but in regions of retrograde winds positive numbers mean downward transport of retrograde angular momentum (see discussion in section \ref{sec:transpmom}). The solid black lines represent the contours of the averaged zonal winds in m$/$s: 4500, 3500, 2500, 1500 and 500. The dashed line represents the contour of averaged zonal winds equal to zero.}
\label{fig:dbvt}
\end{figure}

\subsection{Results}
Below we explore the angular momentum transport in the atmosphere via mean circulation and waves (non-axisymmetric motion). This part was done following the ideas applied to the Venus atmosphere in \cite{2016Mendonca}. The zonally averaged transport of angular momentum is separated into three different contributions: mean circulation [A], stationary waves [B] and transient waves [C]. The mechanical transport from the momentum mixing during atmospheric convection is not explicitly analyzed here because we quantified that this form  of transport has a low impact in the atmospheric circulation compared to the other mechanisms. This conclusion was obtained by comparing the results of the model with and without the momentum mixing representation in the convective adjustment scheme. To analyze the total meridional transport of angular momentum ($M$) in the baseline results we used the following Eq. (\citealt{1992Peixoto}):
\begin{equation}
[\overline{vM}] = \underbrace{[\overline{v}][\overline{M}]}_\text{[A]} + \underbrace{[\overline{v^{\star}}\overline{M^{\star}}]}_\text{[B]} + \underbrace{[\overline{v'M'}]}_\text{[C]}.
\label{eq:mom_trans}
\end{equation}
The disturbances in relation to these two averages are represented by: $M^{\star} = M - [M]$ and $M' = M - \overline{M}$.  To analyze the vertical transport, the variable $v$ is replaced by $w$ in m s$^{-1}$.
For each model cell the angular momentum $M$ is calculated using
\begin{equation}
M = m a \cos \phi (\Omega a \cos \phi + u),
\label{eq:mom}
\end{equation}
where $m$ is the atmospheric mass of a model cell, $u$ the zonal wind, $a$ the planetary radius, $\Omega$ the planetary rotation rate and $\phi$ the latitude. 

When analyzing the angular momentum transport maps we will focus our attention on the main features that are responsible for the formation of the strong winds at low latitudes. In Figs. (\ref{fig:dbht}) and (\ref{fig:dbvt}) we show the maps of angular momentum transport. In these figures, we do not include the maps that show the transport by transient waves because in the reference simulation these waves have relatively low impact in the circulation. The absolute amplitude of the angular momentum transport by the transient waves is not zero but lower than the lowest limit of the range shown in Figs. (\ref{fig:dbht}) and (\ref{fig:dbvt}). The angular momentum transport in the atmosphere is mainly driven by mean circulation and stationary waves. The latter are formed due to the tidally locked state of the planet coupled with the high levels of irradiation. The meridional transport in Fig. \ref{fig:dbht1} is, as expected, consistent with the direction of the mass-stream function shown in Fig. \ref{fig:baselineu}. Angular momentum is transported equatorwards between 10 and 10$^4$ hPa. This transport increases the prograde momentum in that region due to the positive convergence of the angular momentum flux. Above 10 hPa, large direct cells transport angular momentum towards higher latitudes, which reverse again above 1.0 hPa. The stationary waves mostly transport angular momentum (Fig. \ref{fig:dbht2}) towards low latitudes for a large range of pressures (0.3-10$^3$ hPa). This map sustains the idea that atmospheric waves have an important role accelerating prograde winds in the equatorial region. The net meridional transport map shows clearly an equatorial convergence region between 0.3 hPa and pressures below the level 10$^3$ hPa. In some altitude regions the latitudinal region that brings angular momentum towards the equator is shallow. Note that the jet core is located in a region where both mean circulation and wave transport have positive flux convergence.

In Fig. \ref{fig:dbvt}, we show the vertical transport contribution. Notice that the magnitudes in the color-bar in these maps are 3 orders higher than the maps in Fig. \ref{fig:dbht} (meridional transport). Fig. \ref{fig:dbvt1} shows the component associated with the mean circulation. The indirect cells transport angular momentum downwards in a latitudinal shallow region at low latitudes surrounded by regions of upward transport. This downward component is related with the stretching of the jet core vertically. The large-scale direct atmospheric cells transport angular momentum upwards at low latitudes (above 10 hPa) and in the polar regions (downward branches) they transport the angular momentum downwards. The stationary waves created by the strong stellar forcing, transport mostly the angular momentum from the upper atmosphere towards deeper regions in the atmosphere. This transport is related with a positive divergence in the region 100-1000 hPa (acceleration of the momentum in the prograde direction). Upon comparing this map with the zonal wind map in Fig. \ref{fig:baselineu} we can conclude that these stationary waves are decelerating the region associated with the stronger winds. In the net vertical transport map we see larger convergence regions at latitudes 20-50 deg between 20-100 hPa and close to the equator at 1 hPa. Notice that the positive angular momentum transport values at high latitudes in the lower atmosphere are related to downward transport of retrograde momentum, so the boundary between the positive and negative values is not related with convergence of prograde momentum  but with positive divergence. The analysis of the angular momentum transport in the deepest regions of the atmosphere (below 10$^4$hPa) becomes more difficult due to the formation of complex flow structures weakly forced by the direct radiation from the star.

\begin{figure}
\centering
\subfigure[]{\label{fig:db-peri1}\includegraphics[width=0.4\textwidth]{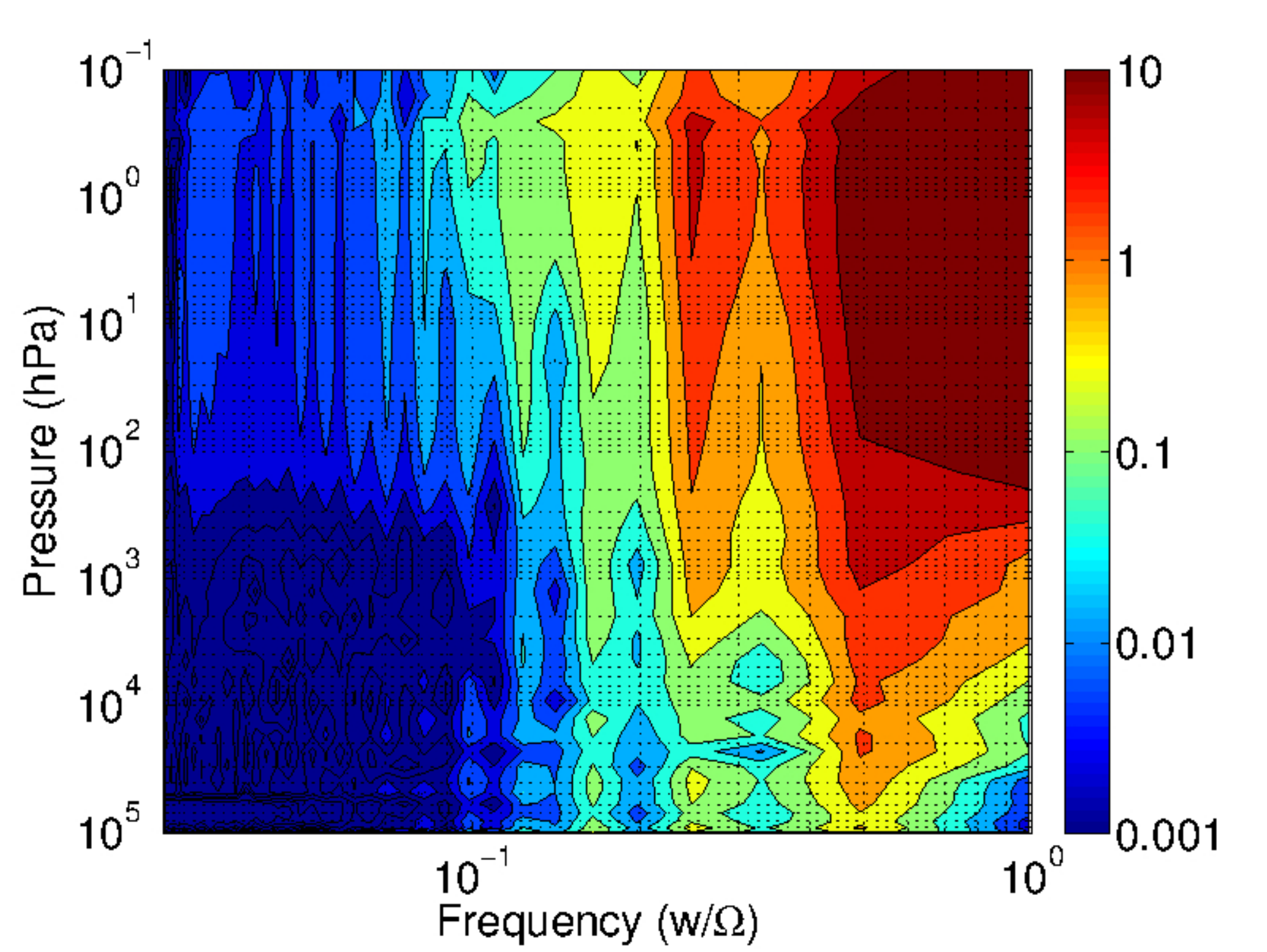}}
\subfigure[]{\label{fig:db-peri2}\includegraphics[width=0.4\textwidth]{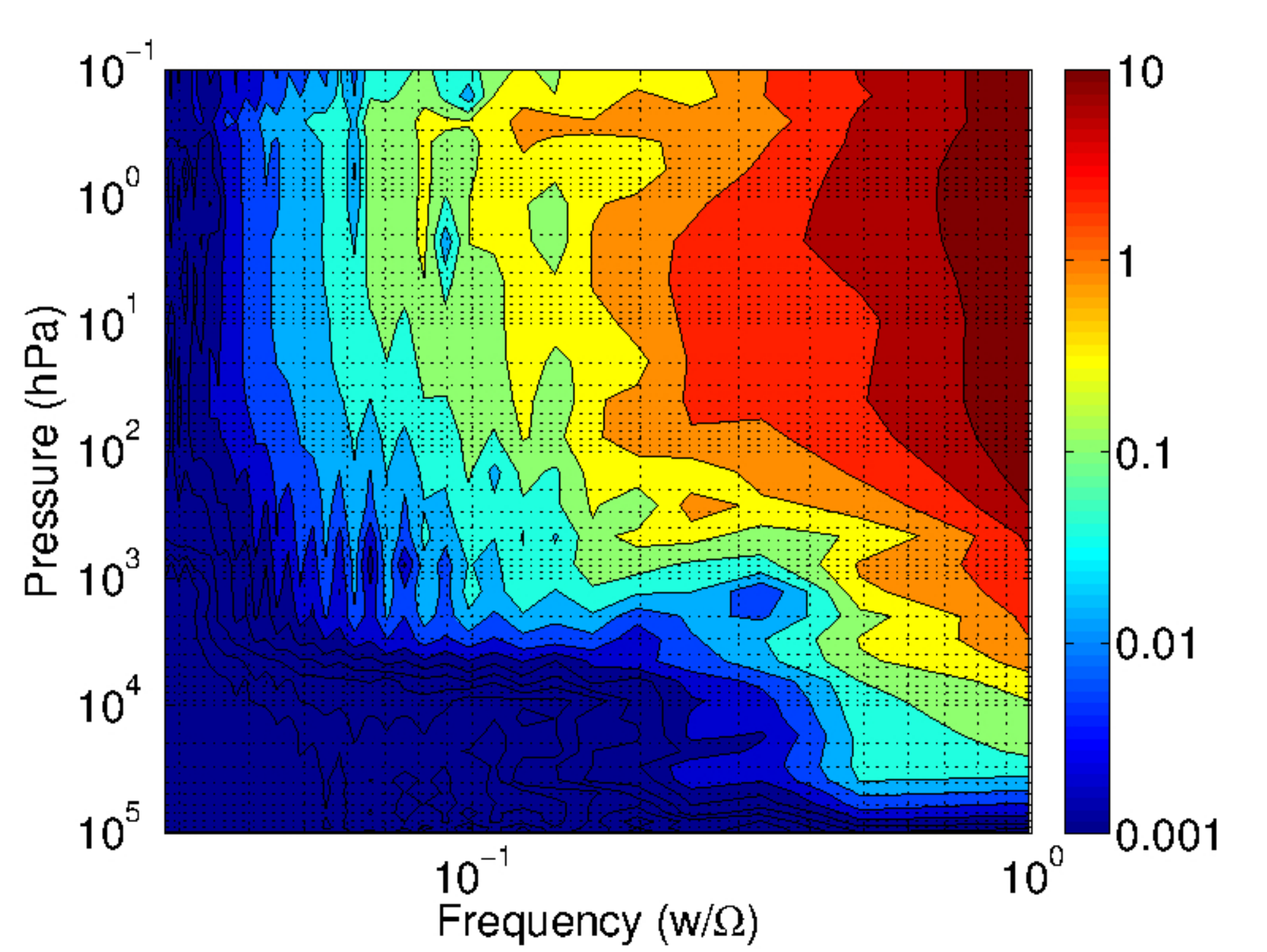}}
\subfigure[]{\label{fig:db-peri3}\includegraphics[width=0.4\textwidth]{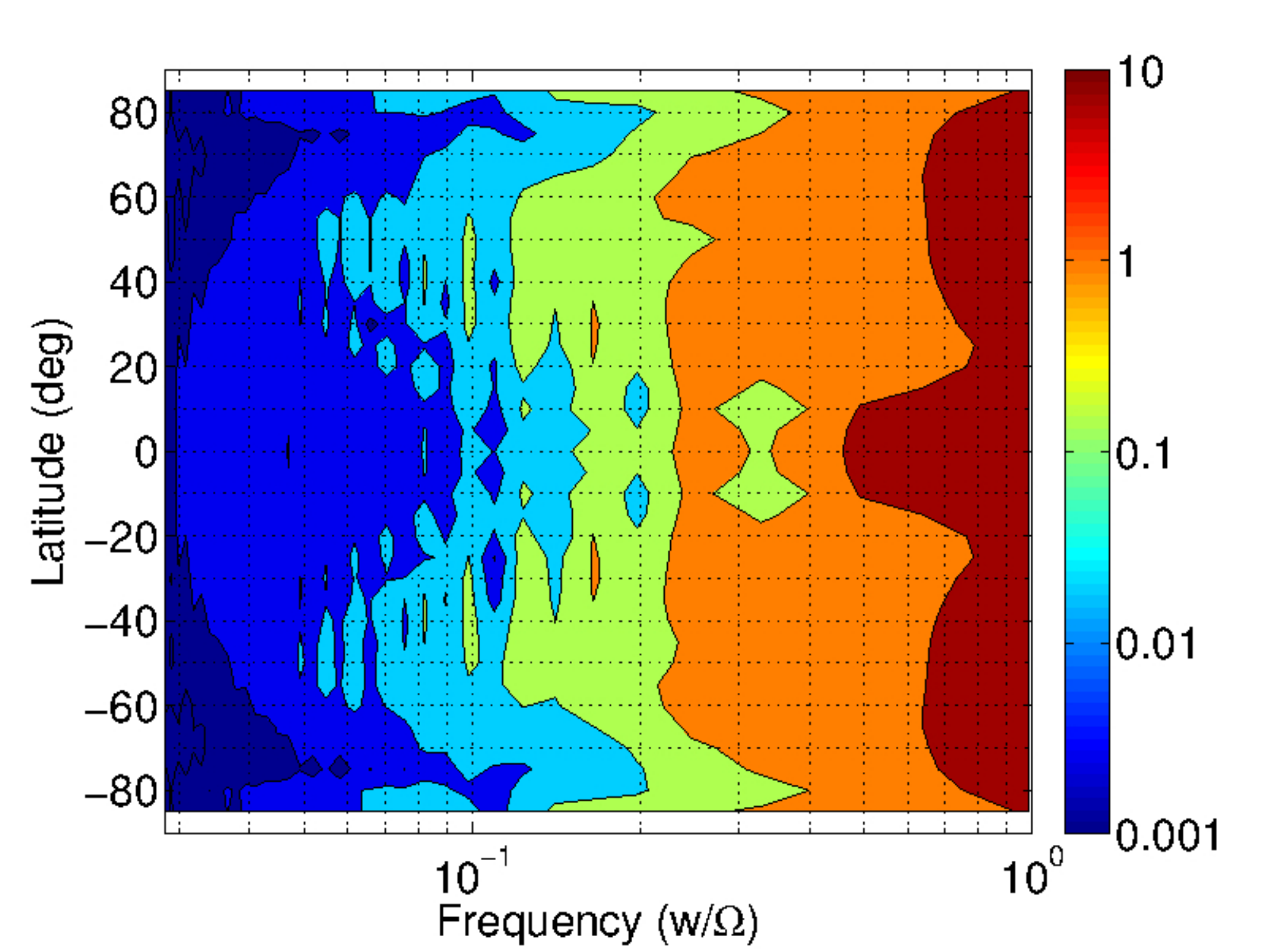}}
\caption{Wave amplitude in the temperature field at the end of the reference simulation. The amplitudes are in Kelvin. The values of the spectra were zonally averaged over 100 Earth days. \textbf{(a)} and \textbf{(b)} show spectra of stationary waves as a function of pressure and frequency for the equator and 50$^{\circ}$N respectively; \textbf{(c)} shows spectra contoured as a function of latitude and frequency at a pressure level of 50 hPa. $\Omega$ is the rotation rate.}
\label{fig:db-peri}
\end{figure}
\begin{figure*}
\begin{center}
\vspace{0.1in}
\includegraphics[width=1.9\columnwidth]{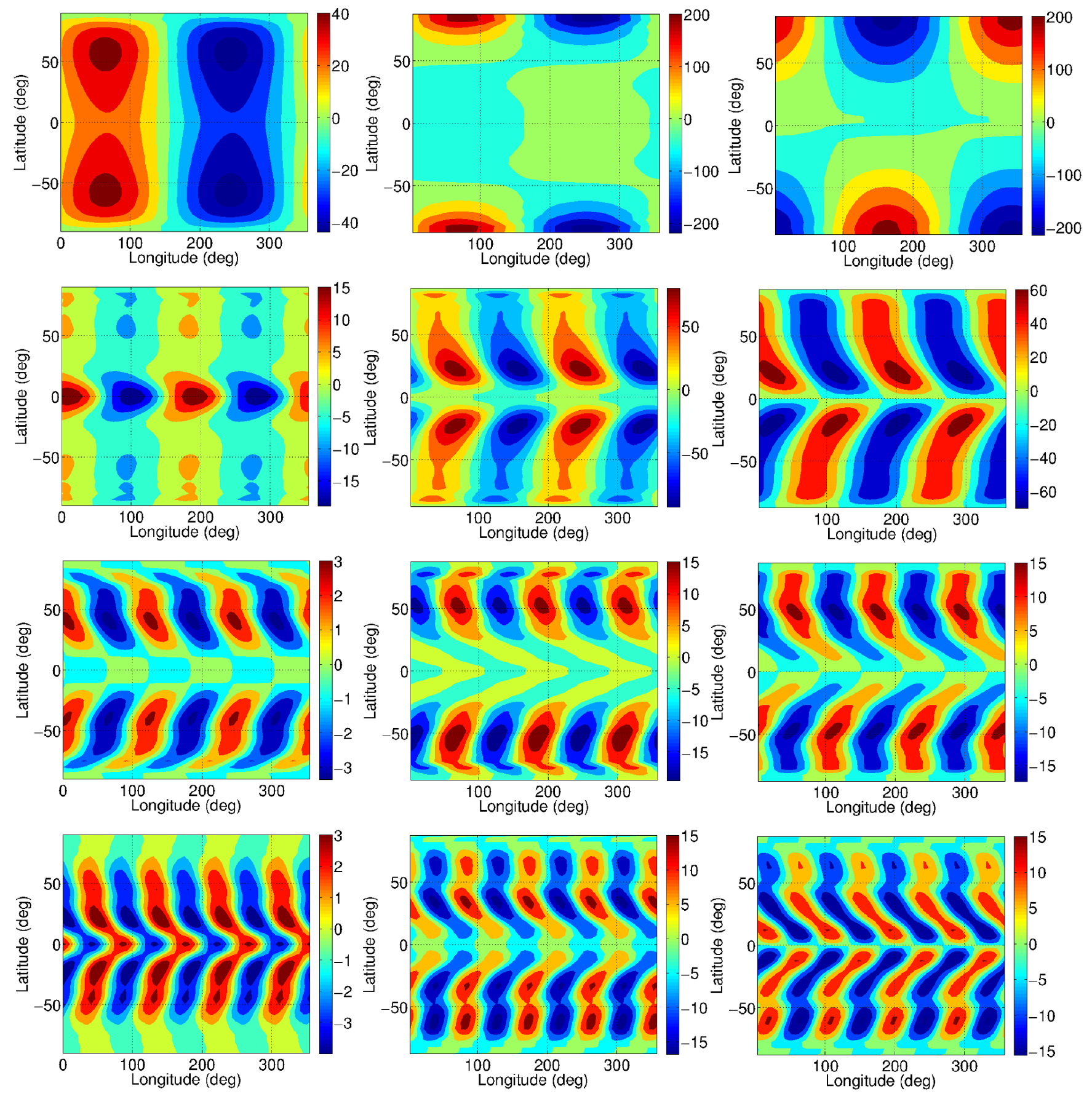}
\end{center}
\caption{These plots correspond to data filtered at different frequencies. The first column corresponds to temperature data, second column zonal winds and the third column meridional winds. The different rows correspond to data filtered at different frequencies: 1, 0.5, 0.33 and 0.25 $w/\Omega$. All the plots are horizontal maps at a pressure level of 100 hPa.}
\label{fig:waves_h}
\end{figure*}

These maps in Figs. (\ref{fig:dbht}) and (\ref{fig:dbvt}) show that the atmospheric circulation is driven mostly by the balance of the  mean circulation and the transport of angular momentum by stationary waves. Despite the larger magnitudes of the vertical angular momentum transport when compared to the horizontal transport, it is the latter that has the most impact in the position of the strong core jet. The direction of the tilting in the atmospheric waves is related to the acceleration/deceleration of the atmospheric flow. The waves that transport most of the angular momentum above 100 hPa at low latitudes are associated with the presence of the semi-diurnal thermal tide (second harmonic of the atmospheric response to the periodic radiative heat forcing from the star). In the next section we will argue that the horizontal tilt of these waves induce an atmospheric acceleration in the prograde direction at low latitudes. 

\section{Wave analysis}
\label{sec:waveanalisis}
In this section, we analyze the main atmospheric wave features present in the reference simulation, which are stationary atmospheric waves as seen in the previous section. We start this analysis by exploring the temperature field data from the last 100 Earth days (20x longer than the rotation period) of the simulation. The temporal resolution is four hours. 

In Fig. \ref{fig:db-peri}, we show the amplitude of the waves in the temperatures field at different pressures, latitudes and frequencies (dimensionless units). These waves travel with the stellar forcing (the planet is tidally locked) and they are related to the angular momentum transport explored in the previous section (they are essential for forming the jet at low latitudes). Figs. \ref{fig:db-peri1} and \ref{fig:db-peri2} show the two  amplitude spectra of the temperature field as a function of pressure for two different locations in latitude (equator and mid-latitude, respectively). The presence of the different harmonics of the thermal tides, such as the diurnal tide (frequency 1) and semi-diurnal tide (frequency 0.5) is clear from these two figures. As we said above, below the pressure level 200 hPa the atmosphere is weakly/non forced by the stellar radiation (see Fig. \ref{fig:fdsg}), so less wave activity is found below this level. However, some wave modes propagate downward, such as $w/\Omega=$ 0.5, 0.25 and 0.125. At mid-latitudes the waves in general propagate deeper until 1000 hPa, which may be related to the region being less turbulent and more transparent to wave propagation. It is also important to remember that very deep in the atmosphere the radiative time scale becomes larger than a sidereal day which diffuses the temperature. The largest amplitudes are located at pressures around 1 hPa. Above this pressure level the waves are mostly absorbed by the atmosphere via radiative damping. In Fig. \ref{fig:db-peri3} we select a pressure level and analyse the amplitude spectra as a function of latitude and frequency. In Fig. \ref{fig:waves_h} we can verify that the amplitudes at low latitudes shown in Fig. \ref{fig:db-peri3} are related to the presence of equatorial Kelvin and Rossby waves. In Fig. \ref{fig:waves_h} one can identify the presence of the different modes of the Rossby waves. These waves are characterized by the symmetry with respect to the equator for the temperature and zonal wind fields and anti-symmetry for the meridional wind field. It is worth noticing that these waves do not reach the equator. In rows two and four of Fig. \ref{fig:waves_h} we see waves clearly peaking in the equatorial region with no meridional velocity component, which is a sign of the presence of Kelvin waves. Despite not being visible in these maps, the equatorial Kelvin waves are also present in rows one and three, however, they are suppressed by the equatorial Rossby waves. The wave structure obtained is a result expected from the theory (see \cite{2014Tsai} for additional discussion). The horizontal structure of these waves is tilted eastwards. It is this tilt that produces the Reynold's stress associated with eddy momentum transport towards low latitudes. Taking into account the impact at low latitudes and the magnitude of the disturbances, roughly five times larger than the other mode, we found that it is the semi-diurnal thermal tide which has the largest impact in the atmospheric circulation and a crucial role in accelerating the jet at low latitudes.

\begin{figure}
\begin{center}
\vspace{0.1in}
\includegraphics[width=0.9\columnwidth]{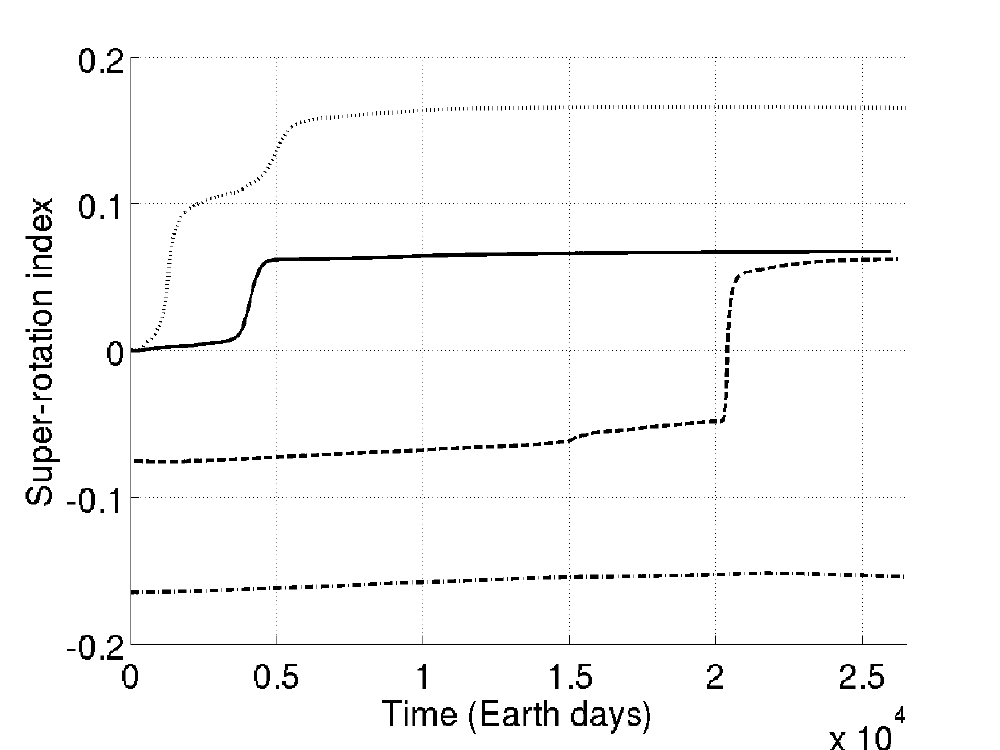}
\end{center}
%\vspace{-0.2in}
\caption{Global super-rotation index for different simulations: reference simulation - solid line; reference simulation but starting with a strong retrograde flow - dashed line; simulation using a rotation rate of 10 Earth days - dotted line; simulation using a rotation rate of 10 Earth days but starting with a strong retrograde flow - dash-dotted line.}
%\vspace{0.1in}
\label{fig:SRind_4ds}
\end{figure}

\section{Multiple equilibrium states?}
\label{sec:MES}
The day/night forcing produces thermal tides which, as described in the previous sections, have an important role in the transport of angular momentum in the atmosphere. In this section we explore the possibility of other equilibrium states using the same model parameters but different initial states. The simulations explored in this section started from rest, reached the equilibrium (end of the spin-up phase) and then the zonal winds were reversed and started again until a new equilibrium was found. The main reason for reversing the zonal winds direction is because we want to force the tilt of the waves which transport angular momentum to change sign as well (i.e., change the direction of the transport). With these conditions, a possible multiple state will be recorded if the mechanisms that maintain the strong retrograde winds are stable for a long period of integration (26500 Earth days in this work). The exploration of this problem can help us to have a better understanding of the atmospheric dynamics at these extreme conditions (i.e., for highly irradiated planets) and on the diversity of possible atmospheric states which can be associated with the history of the planet.  

\begin{figure}
\begin{center}
\vspace{0.1in}
\subfigure[]{\label{fig:u-dt}\includegraphics[width=0.4\textwidth]{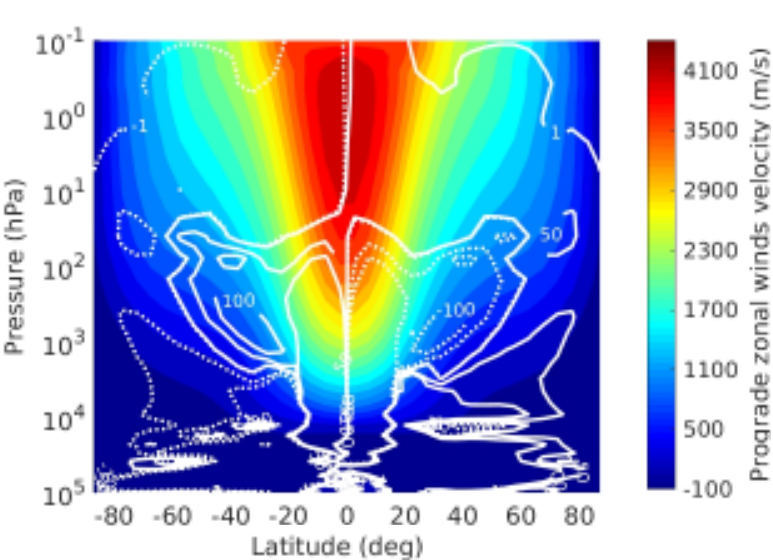}}
\subfigure[]{\label{fig:u-ef}\includegraphics[width=0.4\textwidth]{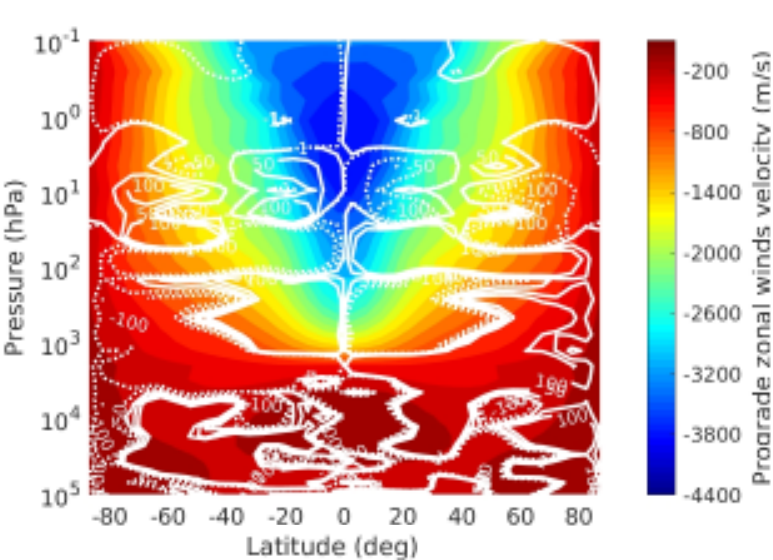}}
\end{center}
%\vspace{-0.2in}
\caption{Averaged zonal winds and mass stream function (in units of $10^{10}$kg/s) from the GCM simulations using a rotation rate of 10 Earth days but starting with different initial conditions:  \textbf{(a)} rest atmosphere and  \textbf{(b)} strong retrograde winds. The dashed lines represent the anti-clockwise circulation and the solid lines the clockwise. The values were zonal and time averaged for 100 Earth days.}
%\vspace{0.1in}
\label{fig:u_ds}
\end{figure}

\begin{figure*}
\centering
\subfigure[]{\label{fig:dtht1}\includegraphics[width=0.45\textwidth]{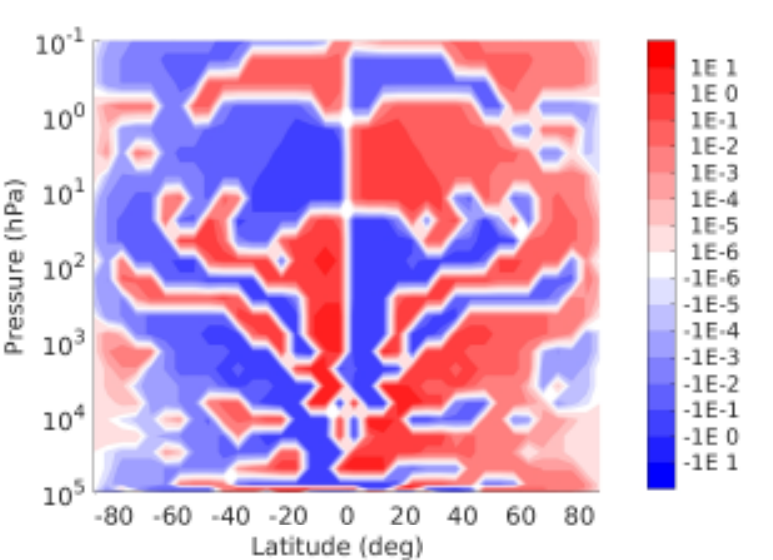}}
\subfigure[]{\label{fig:dtht2}\includegraphics[width=0.45\textwidth]{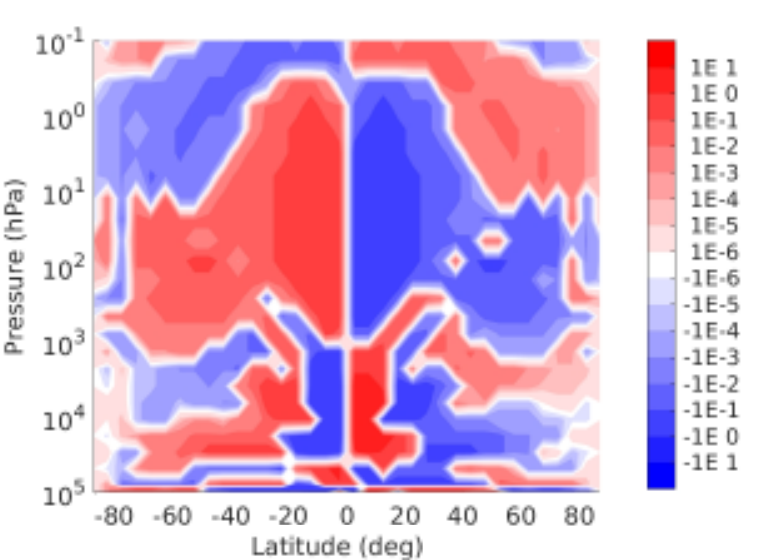}}
\subfigure[]{\label{fig:dtht3}\includegraphics[width=0.45\textwidth]{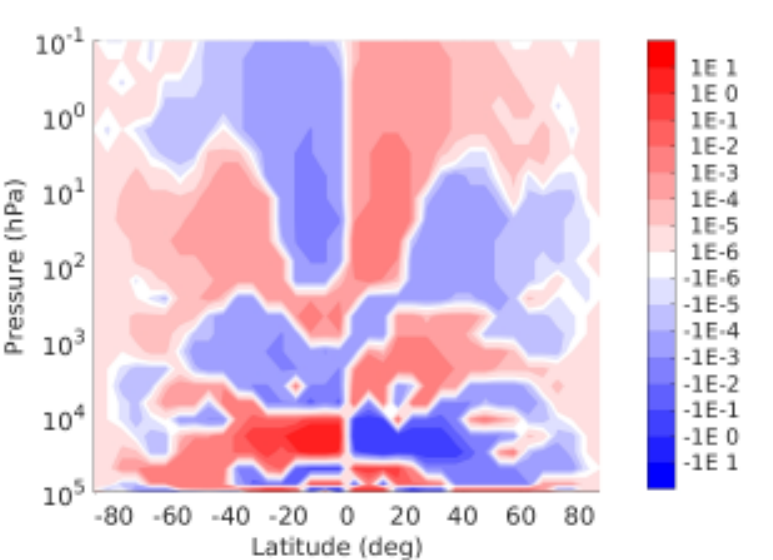}}
\subfigure[]{\label{fig:dtht4}\includegraphics[width=0.45\textwidth]{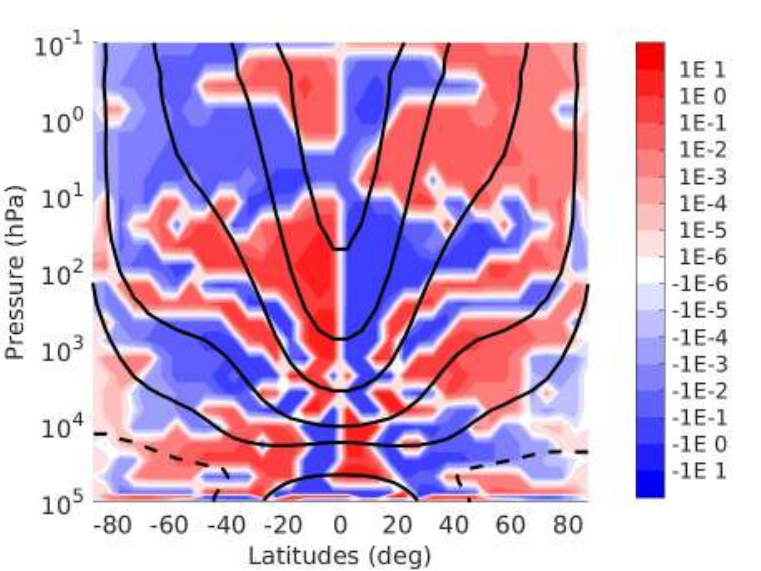}}
\caption{Meridional transport of angular momentum by waves in the GCM simulation using a rotation rate of 10 Earth days and starting from a rest atmosphere. The data used to produce these results correspond to the last 100 Earth days of the simulation. \textbf{(a)} is the horizontal transport by mean ciculation ($[\overline{v}][\overline{M}]$),  \textbf{(b)} by stationary waves ($[\bar{v^{\star}}\bar{M^{\star}}]$),  \textbf{(c)} by transient waves ($[\bar{v'}\bar{M'}]$) and  \textbf{(d)} is the net horizontal transport.  The units of the colour bars are $10^{28}$ kg m$^3$ s$^{-2}$. The solid black lines represent the contours of the averaged zonal winds in m$/$s: 4500, 3500, 2500, 1500 and 500. The dashed line represents the contour of averaged zonal winds equal to zero.}
\label{fig:dtht}
\end{figure*}

\begin{figure*}
\centering
\subfigure{\label{fig:dtvt1}\includegraphics[width=0.45\textwidth]{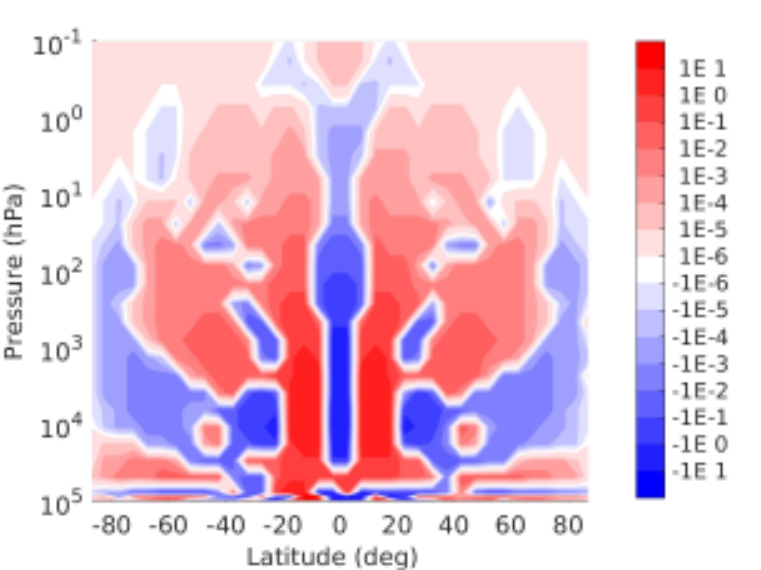}}
\subfigure{\label{fig:dtvt2}\includegraphics[width=0.45\textwidth]{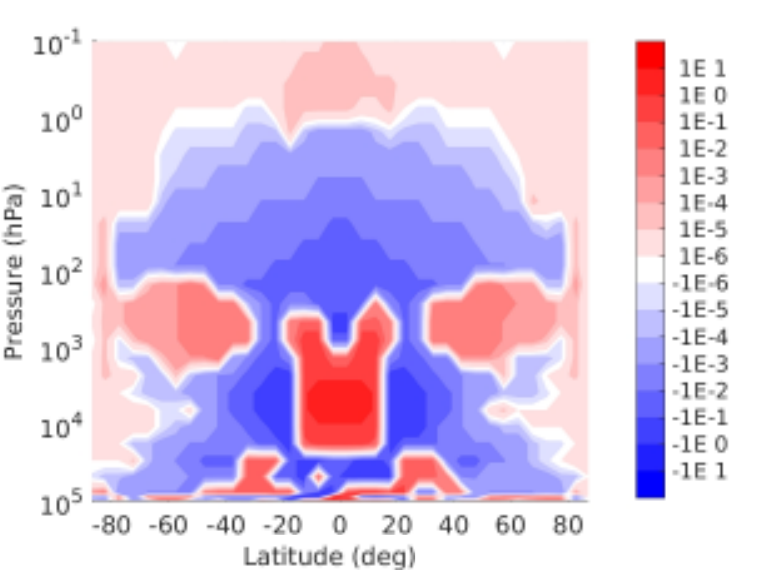}}
\subfigure{\label{fig:dtvt3}\includegraphics[width=0.45\textwidth]{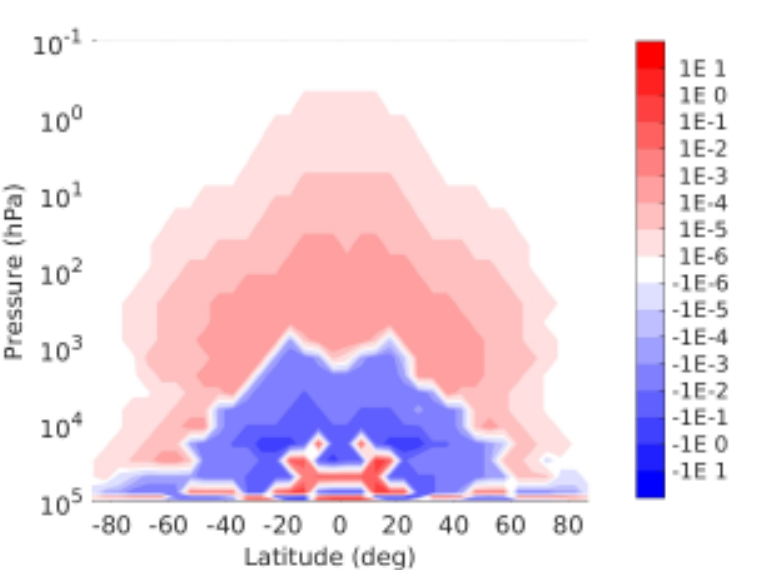}}
\subfigure{\label{fig:dtvt4}\includegraphics[width=0.45\textwidth]{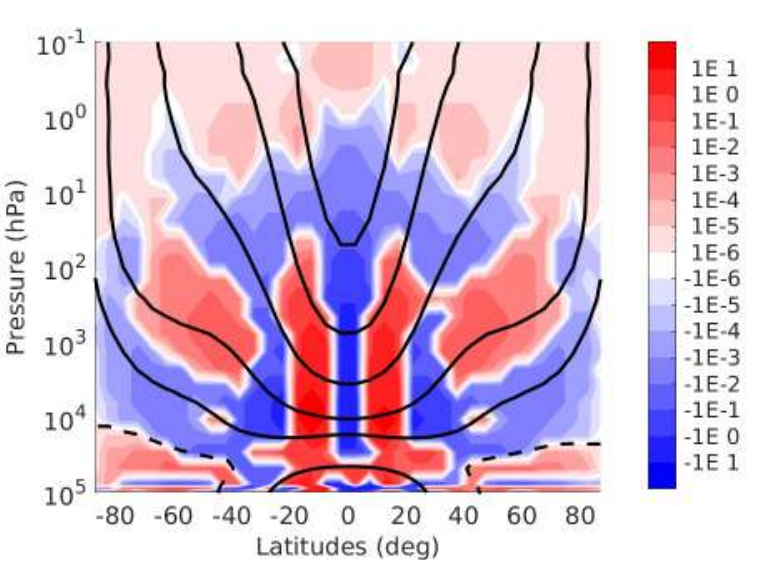}}
\caption{Vertical transport of angular momentum by waves in the GCM simulation using a rotation rate of 10 Earth days and starting from a rest atmosphere. The data used to produce these results correspond to the last 100 Earth days of the simulation. \textbf{(a)} is the vertical transport by mean circulation ($[\overline{w}][\overline{M}]$), \textbf{(b)} by stationary waves ($[\bar{w^{\star}}\bar{M^{\star}}]$),  \textbf{(c)} by transient waves ($[\bar{w'}\bar{M'}]$) and  \textbf{(d)} is the net horizontal transport.  The units of the colour bars are $10^{31}$ kg m$^3$ s$^{-2}$. The solid black lines represent the contours of the averaged zonal winds in m$/$s: 4500, 3500, 2500, 1500 and 500. The dashed line represents the contour of averaged zonal winds equal to zero.}
\label{fig:dtvt}
\end{figure*}

The first experiment explored was the reference simulation. The results obtained in the previous section were used as input to this new experiment, but as mentioned above, the direction of the zonal winds were reversed. Fig. \ref{fig:SRind_4ds} shows the super-rotation index of the atmosphere during the integration of this experiment. As expected, the initial value of the total axial angular momentum is opposite in sign to the value at the end of the simulation in the reference run. The index increases along the integration which is a sign that the atmospheric state with strong retrograde winds is not stable and the main mechanisms transporting angular momentum cannot maintain the retrograde equatorial jet. After a long period of integration the results converge to the ones from the reference simulation. These results also highlight the important need to run the simulations of these massive atmospheres for long periods of time. If the model was stopped before roughly the day 20000 the results would have been an atmosphere with strong retrograde winds but not in equilibrium as we found out later in the simulation. 

Following the results obtained above we decided to run a simulation with a slower rotational rate: ten Earth days. The simulation was started from a rest state and integrated for the same time as the reference simulation. The star-planet distance was kept the same despite being inconsistent with the Kepler law for a tidally locked planet. The reason for this option was to keep the analysis of the results simpler as the differences from the reference simulation will be just due to the change in the rotation rate. However, we did experiments using the correct distance (weaker incoming stellar radiation) and we obtained qualitatively the same results and conclusions as presented below.

The dotted line in Fig. \ref{fig:SRind_4ds} shows that the super-rotation index of the atmosphere reaches an equilibrium. This experiment produces a higher super-rotation index than the reference simulation. Fig. \ref{fig:u-dt} shows a map of the zonal winds and mass stream function for this experiment. The results obtained are similar to the reference simulation, where the main differences are the weaker equatorial jet and the atmospheric cells which extend to higher latitudes. The atmospheric cells extend to higher latitudes due to the weaker Coriolis forces caused by the lower rotation rate. Figs. \ref{fig:dtht} and \ref{fig:dtvt} show the angular momentum transport by mean circulation and atmospheric waves. The meridional transport of angular momentum by stationary waves is stronger at low latitudes than in the reference run. In this case more prograde angular momentum is being transported towards the equatorial region in the pressure region between 1000 hPa and 1 hPa. The effect of transient waves becomes non-negligible and they decelerate the prograde winds in the equatorial region. The transport in the vertical direction is also different from the reference simulation, in this case the wave activity has more weight in the angular momentum transport than the mean circulation above 1000 hPa. As in the meridional projection, the vertical transport by transient waves became more relevant for the experiment with slower rotation rate. In general, these waves transport prograde angular momentum upwards above 1000 hPa accelerating the jet core. It is the growth of the transient waves that enriches the diversity of the atmospheric states. The nature of the transient waves seems to be mostly a manifestation of inertial instabilities in the atmosphere. The inertial instabilities form when the product of the Coriolis frequency and potential vorticity is less than zero (see for example \cite{1987Andrews} for more details). The inertial instabilities are more likely to occur in the equatorial region, and amplify the effects of other waves occurring in the same region (such as the thermal tides) as seen in the wave momentum transport plots.

\begin{figure}
\centering
\subfigure[]{\label{fig:dxvt1}\includegraphics[width=0.45\textwidth]{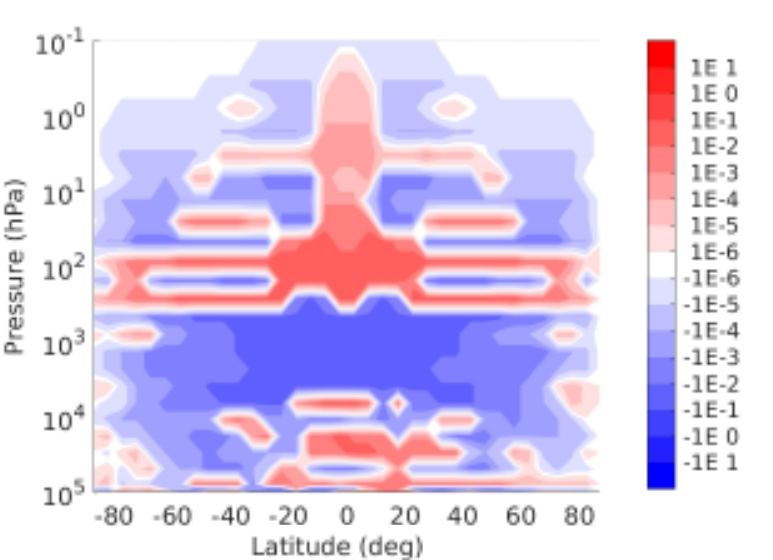}}
\subfigure[]{\label{fig:dxvt2}\includegraphics[width=0.45\textwidth]{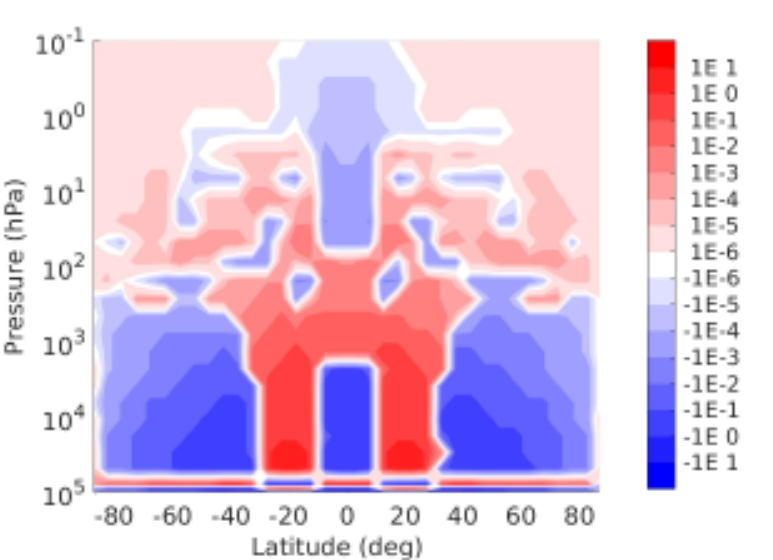}}
\caption{Vertical transport of angular momentum by waves in the GCM simulation using a rotation rate of 10 Earth days and starting with strong winds in the retrograde direction. The data used to produce these results correspond to the last 100 Earth days of the simulation.  \textbf{(a)} shows stationary waves ($[\bar{w^{\star}}\bar{M^{\star}}]$)and  \textbf{(b)} transient waves ($[\bar{w'}\bar{M'}]$).  The units of the colour bars are $10^{31}$ kg m$^3$ s$^{-2}$.}
\label{fig:dxvt}
\end{figure}

After reversing the winds in the zonal direction and integrating for a long time we obtain the winds shown in Fig. \ref{fig:u-ef}. The super-rotation index converges to a value around -0.15, see Fig. \ref{fig:SRind_4ds}. The negative values indicate strong retrograde winds at low latitudes. This experiment is statically stable, and the absolute value of the super-rotation index even increases slightly in the last 5000 Earth days of integration. The distribution of the zonal winds is similar to the previous experiments but with opposite direction and the strong winds do not go deeper than 1000 hPa. The base of the two indirect cells at roughly 20 and 2000 hPa are correlated with two regions of high wave activity as is possible to see in Fig. \ref{fig:dxvt}. This is an indication that stationary waves (thermal tides) are the main responsible for the deeper indirect cell and the transient waves more dominant in the formation of the indirect cell at lower pressures. The mechanism that maintains the strong retrograde jet core at more than 4 km/s is mainly a combination of the angular momentum transport by transient waves and mean circulation.  The upper branches of the upper indirect atmospheric cells transport angular momentum horizontally towards low latitudes and the transient waves above 100 hPa transport angular momentum vertically towards the jet core region. The stationary waves, as in previous experiments, decelerate the jet core by transporting angular momentum from the upper atmosphere to pressures higher than 500 hPa at low latitudes. We note that in this case the colours in Fig. \ref{fig:dxvt} indicate an opposite direction than in previous similar plots because now the zonal winds have a retrograde direction. The weaker dependence of the atmospheric circulation on the rotational effects enhances the impact of the transient waves which makes the circulation at these extreme conditions more sensitive to the initial conditions. These results make the atmospheric flow more like a Venus circulation, however, in Venus the mechanical interaction between the surface and the atmosphere reestablishes a mean preferential direction of the flow (e.g., \citealt{2013Mendonca}). In hot Jupiter planets the interaction of the upper atmosphere with the deep layers in the planets can also have an important role in determining the direction of the strong jet, where MHD drag can force the winds at low latitude to follow the prograde direction. Nevertheless, these results are an example that the diversity of atmospheric circulation at extreme conditions is more rich than previously thought and the current atmospheric state may depend on the planet history.  Observationally this steady state can be validated if the maximum infrared flux in a phase-curve happens consistently after the secondary eclipse.  

These multiple equilibrium states results are physically different from what was found in \cite{2010Thrastarson}. In \cite{2010Thrastarson} a Newtonian relaxation scheme was used to force the temperature in the atmosphere towards a basic state. This very simplified parameterization is used to replace a more computationaly expensive radiative transfer code, however, it depends on two tunable parameters: a three dimensional basic state temperature profile and the Newtonian relaxation time-scale. These two parameters are linked, meaning that if the basic state temperature profile is the radiative equilibrium temperature profile, the time-scale would necessarily have to be the radiative time-scale, however, if the temperature is closer to the one observed, the time-scale needs to be adapted to include e.g., convection phenomena. The latter option is associated with, for example, shorter time-scales in the lower atmosphere than the radiative equilibrium ones. The simulations in \cite{2010Thrastarson} used Newtonian time-scales which represented a very active lower atmosphere (with time-scales shorter than the radiative time-scale values). The very active deep atmosphere consequently makes the upper atmospheric circulation more turbulent, and challenges the dynamical core to resolve the complex flow. Under these conditions the atmospheric flow will depend on initial conditions, dissipation strength and spatial resolution.

\section{Conclusions}
\label{sec:conclu}
In this work we explored the atmospheric circulation of a typically tidally-locked hot Jupiter. The main objective of this work was to study the major mechanisms that are transporting angular momentum and heat in the atmosphere. To explore the main mechanisms that drive the atmosphere we developed a 3D numerical model which contains a quasi-hydrostatic dynamical core, a gray radiative transfer and a convective adjustment scheme. 

A reference simulation was integrated starting from a rest atmosphere until it reached the equilibrium (at 26500 Earth days) using typical parameters that characterize a hot Jupiter planet. The main feature of the atmospheric circulation simulated is a broad jet peaking at the equator and extending from the top of the model domain until 10$^4$ hPa. This jet has a very important role in zonally redistributing heat in the upper atmosphere. We find that the crucial mechanism that transports angular momentum and forms this jet is the coupling between the mean circulation and the presence of the semi-diurnal thermal tide. This tide is excited due to the large contrast of the solar heating in the atmosphere, and its horizontal structure is identified in the results as a coupling between equatorial Rossby and Kelvin waves. This wave structure is tilted eastwards which induces a pressure torque in the atmosphere (Reynold stress) and transports angular momentum towards low latitudes (up-gradient). The convergence of the angular momentum at low latitudes accelerates the winds in the prograde (eastward) direction. This force is balanced by the numerical dissipation and vertical angular momentum transport by the waves. Our results are in line with previous works such as, \cite{2011Showman}, \cite{2014Tsai} and \cite{2017Mayne}. However, we find that the more dominant process driving the circulation is the semi-diurnal tide, which differs from previous works that refer to the diurnal tide (first harmonic) as the main driver. The Fourier analysis and the exploration of the heat transport also allowed us to study for the first time the important thermal eddy-driven cell produced in the equatorial region, and find the first multiple equilibria solutions in 3D simulations of hot Jupiter atmospheres as we discuss below.

Our simulations showed the formation of indirect atmospheric cells at low latitudes. These cells are formed due to large eddy momentum convergence at low latitudes caused by the semi-diurnal tide and they have an important role in coupling the upper with the lower atmosphere. The experiments explored verify that these cells  have an important impact on the atmospheric circulation and climate system. They essentially transport angular momentum and heat from the upper to the lower atmosphere. We find that the indirect cells stretch the jet vertically to deeper regions, and heat is transported from the upper to the deeper atmosphere. The latter works as a ``heat pump'' in the atmosphere, where energy is deposited in deeper layers by the circulation. The impact of parameters such as stellar constant, clouds or gas opacities in the structure of the indirect cells needs to be studied in more detail. 

When slowing the rotation period to a value higher than 5 Earth days, the atmosphere represented by the parameters used in this work can sustain multiple steady state solutions. We found that when the rotation rate is reduced, the transport by stationary and transient waves becomes stronger. The main mechanisms that transport angular momentum become vertical, which coupled with the mean circulation maintains strong zonal winds at low latitudes. These results help us to learn more about the mechanisms that drive the atmospheric circulation in these type of planets. We found that theoretically a planet can sustain a peak in the infrared flux in the observed light curve with a offset consistent with a strong equatorial flow blowing in opposite direction of the rotation of the planet. These multiple solutions need to be taken into account when analyzing observational data, despite the new solution requiring a drastic past history of the planet (e.g., the planet had a different rotation rate in its past). 

We used a dynamical core with simplified but comprehensive physics, which facilitates the interpretation of the model results, while being less computational expensive and providing a basic framework for testing theories of the general circulation. A much larger parameter space needs to be explored, including evaluating the impact of clouds which continues to be poorly studied. A better understanding of the vast diversity of atmospheric circulations and their mechanisms will be a crucial step to characterize these extreme climate conditions. 

\section*{Acknowledgements}
J.M.M. thanks the Met Office for the use of the HadAM3 dynamical core in this work, and Profs. Peter Read and Kevin Heng, and Dr. Shang-Min Tsai for helpful discussions. J.M.M. acknowledges financial support from the VILLUM Foundation YIP Program and the Carlsberg Foundation Distinguished Associate Professor Fellowship. 

%%%%%%%%%%%%%%%%%%%%%%%%%%%%%%%%%%%%%%%%%%%%%%%%%%

%%%%%%%%%%%%%%%%%%%% REFERENCES %%%%%%%%%%%%%%%%%%
\bibliographystyle{mnras}
%\bibliography{mybibfile} % if your bibtex file is called example.bib

%%%%%%%%%%%%%%%%%%%%%%%%%%%%%%%%%%%%%%%%%%%%%%%%%%

% Don't change these lines
\bsp	% typesetting comment
\label{lastpage}
\end{document}